\documentclass[10pt]{article}

\usepackage{a4wide}
\usepackage{amsmath}
\usepackage{amssymb}
\usepackage{mathabx}
\usepackage{amsthm}
\usepackage{mathtools}
\usepackage[mathscr]{euscript}
\usepackage{graphicx}
\usepackage[margin=10pt,font=small,labelfont=bf,labelsep=endash]{caption}
\usepackage{subcaption}
\usepackage{threeparttable}
\usepackage{booktabs}
\usepackage[bookmarks=true,bookmarksopen=true,colorlinks=true,breaklinks=true,linkcolor=black,citecolor=black]{hyperref}

\newcommand{\ds}{\,\mathrm{d}s}
\newcommand{\dE}{\,\mathrm{d}E}
\newcommand{\dz}{\,\mathrm{d}z}
\newcommand{\drho}{\,\mathrm{d}\rho}

\newtheorem{theorem}{Theorem}[section]
\newtheorem{definition}[theorem]{Definition}

\newtheorem{remark}[theorem]{Remark}

\numberwithin{equation}{section}

\newcommand {\R} {\mathbb R}

\newcommand{\Eqref}[1]{Eq.~\eqref{#1}}
\newcommand{\Eqsref}[1]{Eqs.~\eqref{#1}}
\newcommand{\Sectionref}[1]{Section~\ref{#1}}

\newcommand{\Figref}[1]{Figure~\ref{#1}}
\newcommand{\Remarkref}[1]{Remark~\ref{#1}}
\newcommand{\Figsref}[1]{Figures~\ref{#1}}
\newcommand{\Tableref}[1]{Table~\ref{#1}}

\begin{document}

\title{On Axisymmetric and Stationary Solutions
of the Self-Gravitating Vlasov System}
\author{
Ellery Ames
\thanks{Department of Mathematical Sciences, Chalmers University of Technology and University of Gothenburg}
\footnote{ames@chalmers.se}
\and
H\aa kan Andr\'easson\footnotemark[1]
\footnote{hand@chalmers.se}
\and
Anders Logg\footnotemark[1]
\footnote{logg@chalmers.se}
}
\date{\today}
\maketitle

\begin{abstract}
Axisymmetric and stationary solutions are constructed to the Einstein--Vlasov and Vlasov--Poisson systems. These solutions are constructed numerically, using finite element methods and a fixed-point iteration in which the total mass is fixed at each step. A variety of axisymmetric stationary solutions are exhibited, including solutions with toroidal, disk-like, spindle-like, and composite spatial density configurations, as are solutions with non-vanishing net angular momentum. In the case of toroidal solutions, we show for the first time, solutions of the Einstein--Vlasov system which contain ergoregions.
\end{abstract}


\section{Introduction}
While the self-gravitating Vlasov system has proven to be a useful model in astrophysics, and serves as a well-defined matter model in general relativity, the space of axisymmetric solutions is still poorly understood. These models are well-studied under the restriction to spherical symmetry; see \cite{binney2011galactic,Rein:ud} and references therein for the Vlasov--Poisson case, and \cite{Andreasson:2011dza} for a review in the Einstein--Vlasov case. In going beyond spherically symmetry however, the equations become much more complicated and few mathematical or numerical results have been established.

The purpose of this article is to construct solutions to the axisymmetric self-gravitating Vlasov system via a numerical method.
We start with an ansatz that the Vlasov distribution depends on the phase space coordinates only through a function of the two classical integrals of motion, and solve for the spatial density and gravitational potentials --- the Newtonian potential, or in the Einstein case, the metric fields. The solutions we obtain are thus guaranteed to be fully self-consistent.
In this paper we investigate solutions which are obtained from ansatz functions of simple form, as well as compositions thereof.

Shapiro and Teukolsky have also studied the self-gravitating Vlasov system, and numerically construct axisymmetric solutions to the Vlasov--Poisson \cite{Shapiro:1992ei} and the Einstein--Vlasov \cite{Shapiro:1993gb,Shapiro:1993hi} systems. Notably, they obtain solutions which are far from spherically symmetric in the relativistic case. 
In contrast to the Vlasov-Poisson system, where rigorous existence of axisymmetric solutions which are not necessarily close to spherically symmetric are known \cite{Rein:2003cg}, the only rigorous existence results for axisymmetric solutions to the Einstein-Vlasov system are for solutions that are perturbed off of spherically symmetric Newtonian solutions. These results are due to Andr\'easson et al. in \cite{Andreasson:2011hg} for the static case, and in \cite{Andreasson:ch} for the stationary case. 
It remains an interesting open question to prove the existence of solutions which are far from spherically symmetric, and it is hoped that the numerics employed here will eventually help guide a suitable method of proof.

The present work validates, but also extends the above mentioned work of Shapiro and Teukolsky. In particular we are able to generate highly relativistic configurations which contain ergoregions. This is the first instance, to the authors' knowledge, of such results in the Einstein--Vlasov literature. Ansatz functions can be easily and rapidly implemented in our code allowing for exploration of the vast solution space of the self-gravitating Vlasov system. In the present paper we illustrate several different choices of ansatz which generate toroidal, disk-like, and, spindle-like solutions, as well as composite solutions formed from the sum of multiple ansatz functions.

\Sectionref{sec:EquationsAndSetUp} of the paper contains a presentation of the equations, including both the Vlasov--Poisson and the Einstein--Vlasov systems, the boundary conditions, and the physical characteristics of the solutions which are monitored in the numerical simulations. We present in  \Sectionref{sec:NumericalMethods} the numerical method which is used in solving these equations. \Sectionref{sec:Results} is devoted to numerical results, where we present generalized polytropic solutions (\Sectionref{sec:PolytropicAnsatzSolutions}), relativistic toroidal solutions with ergoregions  (\Sectionref{sec:ThinTorusSolutions}), disk-like solutions  (\Sectionref{sec:Disks}), spindle-like solutions  (\Sectionref{sec:SpindleSolutions}), and composite objects  (\Sectionref{sec:CompositeObjects}).

\section{The Axisymmetric Equations}
\label{sec:EquationsAndSetUp}
\subsection{Self-Gravitating Vlasov Matter}
Self-gravitating Vlasov matter models a large collection of particles which do not interact pair-wise via collisions, but only through the collective gravitational field generated by the particles. For this reason it is sometimes called \emph{collisionless matter}. The model is statistical in that the matter is described by a density function $f: P \to [0,\infty[$. In Newtonian theory and in three space dimensions $f = f(t,x,p)$ (i.e. $P = \mathbb R \times \mathbb R^{6}$).
In the framework of general relativity the density function is defined on a subset of the tangent bundle of a  time-oriented Lorentzian manifold $(M,g)$, called the mass-shell. For particles of mass $m$, the mass-shell $P$ is defined as the set of all future pointing time-like vectors $p$ of square length $g(p,p) = -m^2$. The model is simplified by taking all particles to have the same mass, which we set to one.
Below we introduce the coupled self-gravitating Vlasov matter systems when gravity is modeled using Newton's equations, and with Einstein's equations. More thorough introductions to these systems can be found in \cite{binney2011galactic,Rein:ud} (Vlasov--Poisson case), and \cite{Andreasson:2011dza} (Einstein--Vlasov case).

\subsection{The Axisymmetric Vlasov--Poisson System}
\label{sec:VlasovPoissonSystem}
The Vlasov--Poisson system for the distribution function $f$ introduced above and the potential $U: \mathbb R\times \mathbb R^3 \to \mathbb R$ reads

\begin{align}
\label{eq:Vlasov}
\partial_t f(t,x,p) + p \cdot \nabla_x f(t,x,p) + \nabla_x U(t,x) \cdot \nabla_p f(t,x,p) = 0, \\
\label{eq:Poisson}
\Delta U(t,x)  =  4 \pi w_f(t,x), \quad \lim_{|x| \to \infty} U(t,x) = 0, \\
\label{eq:VPDensityDef}
w_f(t,x) = \int_{\mathbb R^3} f(t,x,p) \mathrm{d}^3p.
\end{align}
It is conventional to denote the spatial density by the letter $\rho$. However, since we use $\rho$ to denote the cylindrical radial coordinate below, we instead choose $w$ here and below. Our aim is to numerically compute static solutions $f_0(x,p), U_0(x)$ to \Eqsref{eq:Vlasov}--\eqref{eq:Poisson} under the assumption of axisymmetry.

The above equations are reduced to a semilinear elliptic equation for the potential $U_0(x)$ by making an ansatz of the form
\begin{equation}
\label{eq:AnsatzForm}
f_0 = K\Phi(E, L_z),
\end{equation}
where $K$ is a normalization constant to be determined, and where $E = \frac 12 p^2 + U_0(x)$ is the particle energy and $L_z = x_1p_2 - x_2 p_1$ is the particle angular momentum about the axis of symmetry, which we take to be the $(x_3 \equiv z)$-axis. Since these quantities are conserved under the particle motion, the Vlasov equation \eqref{eq:Vlasov} is automatically satisfied. The density becomes a functional of the potential $w_0(U_0) := w_{f = f_0}$ leading to the system
\begin{equation}
\label{eq:SemilinearPoisson}
\Delta U_0(x) = 4 \pi w_0(U_0(x)), \quad  \lim_{|x| \to \infty} U_0(x) = 0.
\end{equation}

Let $(\rho, z, \varphi)$ denote the usual axial coordinates. We may write the momentum-space integral in \Eqref{eq:VPDensityDef} in terms of the coordinates $p_{\varphi}$, and  $(p_\rho, p_z) = (p_m \cos \eta, p_m \sin \eta)$, with volume element $\mathrm{d}^3p = p_m \mathrm{d}p_m \mathrm{d}p_{\varphi} \mathrm{d}\eta$. In terms of these coordinates
\[
E = \frac 12 (p_{\varphi}^2 + p_m^2) + U_0(\rho, z),
\quad \text{and} \quad
L_z = \rho p_{\varphi}.
\]
Changing the integration variables to $(E, p_\varphi, \eta)$, and using the Jacobian determinant $1/(p_m)$ we obtain,
\begin{align}
\label{eq:VPDensityIntegral}
\begin{split}
w_0(U_0(\rho,z))
&= \int_0^{2\pi} \mathrm{d}\eta \int_{U_0(\rho,z)}^\infty \int_{-\overline p_{\varphi}}^{\overline p_{\varphi}} K \Phi(E, \rho p_\varphi) \,\mathrm{d}p_\varphi \dE,   \\
&= 2\pi \int_{U_0(\rho,z)}^\infty \int_{-\overline p_{\varphi}}^{\overline p_{\varphi}} K \Phi(E, \rho p_\varphi)  \,\mathrm{d}p_\varphi \dE.
\end{split}
\end{align}
where $\overline p_\varphi := \sqrt{2(E - U_0)}$.


\subsection{The Axisymmetric Einstein--Vlasov System}
\label{sec:EinsteinVlasovSystem}
The Einstein--Vlasov system consists of the coupled equations for the metric tensor $g$ and distribution function $f$, which in arbitrary coordinates and geometric units ($G = c = 1$) reads
\[ \mathrm{Ric}(g)_{ij} - \frac 12 R(g) g_{ij} = 8 \pi T(g,f)_{ij}, \quad p^i \partial_{x^i} f - \Gamma^{k}_{ij}(g) p^i p^j \partial_{p^k} f = 0. \]
Here $\mathrm{Ric}(g), R(g)$ are the Ricci tensor and Ricci scalar of the metric $g$, $\Gamma^{k}_{ij}(g)$ are the Christoffel symbols of the metric $g$, and $T_{ij}(g,f)$ is the energy momentum tensor associated with the Vlasov matter.

For the stationary axisymmetric spacetimes considered in this paper the metric can be written in axial coordinates $(t, \rho, z, \varphi)$ (following  \cite{Bardeen:1973ux}) as
\begin{equation}
\label{eq:Metric}
g = - e^{2 \nu} dt^2 + e^{2 \mu} d\rho^2 + e^{2 \mu}dz^2 + \rho^2 B^2 e^{-2 \nu} (d\varphi- \omega dt)^2,
\end{equation}
where the metric fields $\nu, \mu, B, \omega$ depend only on the coordinates $\rho, z$. Note that $\rho = 0$ is the axis of symmetry, and that $(\rho, z)$ are cylindrical coordinates at infinity in the sense that in the appropriate limit $\rho$ is the radius of the symmetry group orbits. The metric field $\omega$ identically vanishes for solutions with no net rotation.

It is useful, as in Andr\'easson et al. \cite{Andreasson:ch}, to introduce the following frame
\begin{equation}
\label{eq:VBasis}
 v^0 = e^\nu p^0, \quad v^1 = e^\mu p^1, \quad v^2 = e^\mu p^2, \quad v^3 = \rho B e^{-\nu} ( p^3  - \omega p^0).
\end{equation}
The time-independent energy momentum tensor can then be written as
\begin{equation}
\label{eq:EMTensorVCoords}
T_{ij}(\rho, z,\varphi) = \int_{\mathbb R^3} p_i p_j f_0(\rho, z,\varphi,v^1,v^2,v^3) \frac{\mathrm{d}^3 v}{\sqrt{1 + |v|^2}},
\end{equation}
where $p_i$ is obtained from \Eqref{eq:VBasis} and the relation $p_i = g_{ij}p^j$, and $\mathrm{d}^3v := \mathrm{d}v^1 \mathrm{d}v^2 \mathrm{d}v^3$. In particular, this choice allows one to consider solutions which contain ergoregions; for a more detailed discussion of the issues see \cite{Andreasson:ch}. Moreover, $f$ is taken to be a function on the forward mass-shell, expressed in the $v$-basis as the positive root of $(v^0)^2 =  1 + |v|^2$, which represents that all particles move forward in time.

As in the Vlasov--Poisson case, we make an ansatz that the distribution function depends on position and momentum through the particle energy $E$ and angular momentum $L_z$ about the axis, $f_0 = K \Phi(E, L_z)$. These quantities, for which we have the expressions
\begin{align*}
E   	&= -g(\partial_t, p^i) = e^{2 \nu} p^0 + \omega (\rho B)^2 e^{-2\nu}  (p^3 - \omega p^0), \\
L_z 	&= g(\partial_\varphi, p^i) = (\rho B)^2 e^{-2 \nu} ( p^3 - \omega p^0),
\end{align*}
are constant along the geodesic flow, and hence the Vlasov equation is satisfied.
In terms of the frame \Eqref{eq:VBasis} we compute
\[  L_z =  \rho B e^{-\nu} v^3 =: \rho s \]
and
\[ E = e^\nu \sqrt{1 + |v|^2} + \omega L_z =: h + \omega \rho s. \]

With these definitions the energy momentum tensor components can be seen to become integral expressions in the metric fields. It is convenient to introduce the following combinations of the components of the energy momentum tensor, $\Phi_{00}, \Phi_{11},\Phi_{33},\Phi_{03}$, and perform the integration over $h$ and $s$. Let
\begin{align}
\label{eq:Phi00_expression}
\nonumber
\Phi_{00} 	&=  e^{2\mu - 2 \nu} T_{00} \\
		&= \frac{2\pi}{B} e^{2\mu - 2\nu} \int_{e^\nu}^\infty \int_{-\overline s}^{\overline s} E(h,s)^2 K\Phi(E(h,s), \rho s) \ds \,\mathrm{d}h, \\
\nonumber
\Phi_{11} 	&= T_{\rho \rho} + T_{zz}  \\
		&= \frac{2\pi}{B^3} e^{2\mu + 2 \nu} \int_{e^\nu}^\infty \int_{-\overline s}^{\overline s} \left( \overline s^2 - s^2 \right) K\Phi(E(h,s), \rho s) \ds \,\mathrm{d}h, \\
\label{eq:Phi33_expression}
\nonumber
\Phi_{33} 	&= (\rho B)^{-2} e^{2\mu + 2 \nu} T_{\varphi \varphi} \\
		&=\frac{2\pi}{B^3} e^{2\mu + 2 \nu} \int_{e^\nu}^\infty \int_{-\overline s}^{\overline s} s^2 K\Phi(E(h,s), \rho s) \ds \,\mathrm{d}h, \\
\label{eq:Phi03_expression}
\nonumber
\Phi_{03} 	&= e^{2\mu + 2 \nu} T_{0 \varphi} \\
		&= - 2\pi \rho B^{-1} e^{2\mu + 2 \nu}\int_{e^\nu}^\infty \int_{-\overline s}^{\overline s} s E(h,s) K\Phi(E(h,s), \rho s) \ds \,\mathrm{d}h,
\end{align}
where
\begin{equation}
\label{eq:SlDef}
\overline s := B e^{-\nu} \sqrt{e^{-2\nu} h^2 -1} .
\end{equation}
These expressions can be seen to agree with those obtained by Andr\'easson et al. in \cite{Andreasson:ch} if we note that here we use the energy $E$ rather than $\eta := E-1$ used in that paper, and in addition we set their parameter $\gamma$ to one. The definition of $\Phi_{33}$ used here also contains an additional scaling factor of $(\rho B)^{-2}$ compared to that in \cite{Andreasson:ch}.

As a result of the ansatz \eqref{eq:AnsatzForm} and the above definitions, the Einstein--Vlasov system in this case reduces to the following system of semi-linear elliptic equations for the metric fields
\begin{align}
\label{eq:EinsteEqNU}
\Delta \nu & =
4 \pi \left(
\Phi_{00} + \Phi_{11}
+ \left( 1 + (\rho B)^2 e^{-4 \nu} \omega^2 \right)\Phi_{33}
+ 2 e^{-4 \nu} \omega \Phi_{03} \right)  \\\nonumber
& - \frac 1B \nabla B \cdot \nabla \nu
+ \frac 12 e^{-4\nu} (\rho B)^2 \nabla \omega \cdot \nabla \omega,  \\
\label{eq:EinsteEqBB}
\Delta B & = 8 \pi B \Phi_{11}  - \frac 1\rho \nabla \rho \cdot \nabla B, \\
\label{eq:EinsteEqMU}
\Delta \mu & =
- 4 \pi \left(
\Phi_{00} + \Phi_{11}
+ \left((\rho B)^2 e^{-4 \nu} \omega^2 - 1 \right)\Phi_{33}
+ 2 e^{-4 \nu} \omega \Phi_{03} \right)  \\ \nonumber
& + \frac 1B \nabla B \cdot \nabla \nu - \nabla \nu \cdot \nabla \nu
+ \frac 1\rho \nabla \rho \cdot \nabla \mu + \frac 1\rho \nabla \rho \cdot \nabla \nu
+ \frac 14 e^{-4\nu} (\rho B)^2 \nabla \omega \cdot \nabla \omega ,  \\
\label{eq:EinsteEqWW}
\Delta \omega & = \frac{16 \pi}{(\rho B)^2} \left( \Phi_{03} + (\rho B)^2 \omega \Phi_{33} \right)
- \frac 3B  \nabla B \cdot \nabla \omega + 4 \nabla \nu \cdot \nabla \omega
- \frac 2\rho \nabla \rho \cdot \nabla \omega.
\end{align}
Here $\Delta, \nabla$ are respectively the laplacian and gradient in cartesian coordinates, and $a \cdot b$ represents the scalar product with respect to the Euclidean metric of $a$ and $b$.

It is also sometimes convenient to use the field $\xi := \nu + \mu$. The Einstein equations imply two equations involving $\partial_\rho \xi$ and $\partial_z \xi$ (see \cite{Andreasson:ch}), and from these we can derive an equation\footnote{\Eqref{eq:XIequation} corrects a minor typo in Equation 2.13 of \cite{Andreasson:ch}.} only in $\partial_\rho \xi$:
\begin{align}
\label{eq:XIequation}
\begin{split}
\left( (B + \rho B_\rho)^2 + (\rho B_z)^2 \right) \partial_\rho \xi
& = (B + \rho B_\rho)
	\left(
		B_\rho + \frac \rho2 (B_{\rho \rho} - B_{zz})
	\right)
+ \rho B_z (B_z + \rho B_{z \rho}) \\
& + (B + \rho B_\rho) \rho B \left( \nu_\rho^2 - \nu_z^2\right)
+ 2 \rho^2 B B_z \nu_\rho \nu_z \\
& - (B + \rho B_\rho)
	\left(
		\rho^3 B^3 e^{-4 \nu} (\omega_\rho^2 - \omega_z^2 )
	\right)
+ \frac 12 \rho B_z (\rho B)^3 e^{-4 \nu} \omega_\rho \omega_z.
\end{split}
\end{align}
One may replace \Eqref{eq:EinsteEqMU} for $\mu$ in the Einstein system with the above equation for $\xi$. The advantage in some cases comes from the fact that solving \Eqref{eq:XIequation} requires only an integration in the radial coordinate.

\subsection{Boundary Conditions}
In order to solve the equations \eqref{eq:SemilinearPoisson} in the Vlasov--Poisson case, and  \Eqsref{eq:EinsteEqNU}--\eqref{eq:EinsteEqWW} in the Einstein--Vlasov case we must impose boundary conditions.
For the Vlasov--Poisson system we prescribe
\begin{equation}
\label{eq:VlasovPoissonBCInfinity}
\lim_{|(\rho,z)| \to \infty} r U_0(\rho, z) = - M,
\end{equation}
where $r = |(\rho,z)|$ and $M$ is the total mass of the particles given by
\begin{equation}
M = 2 \pi \int_{\mathbb R^2} w_0(\rho,z) \rho\, \drho \dz.
\end{equation}
This boundary condition is exact in the spherically symmetric case, and serves as a leading order approximation in axisymmetry. However, the error in the boundary condition can be reduced by taking a computational domain which is large compared to the matter support.

For the Einstein--Vlasov system we seek solutions which are asymptotically flat, from which it follows \cite{Bardeen:1973ux,Ansorg:2008jr} that
\[ \nu, \mu, \omega \to 0 \quad \text{and} \quad B \to 1 \quad \text{as} \quad r = |(\rho,z)| \to \infty, \]
and
\begin{align}
\label{eq:EinstBCInfinity}
\nu & = - \mathcal M / r + O(r^{-2}),  &
\mu & = \mathcal M / r + O(r^{-2}),  &
\omega & = 2 \mathcal J / r^3 + O(r^{-4}), &
B = 1 + O(r^{-2}),
\end{align}
where $\mathcal M$ is the total mass of the system, given by \Eqref{eq:TotalMassEnergy} below, and $\mathcal J$ is the total angular momentum computed via \Eqref{eq:TotalAngularMomentum}.
In addition we require that the metric be locally flat at the axis, which implies
\begin{equation}
\label{eq:EinstBCAxis}
\nu(0, z) + \mu(0,z) = \ln B(0,z)
\end{equation}
for all $z$ in the solution domain.

\subsection{Solution Characteristics}
\label{sec:SolutionCharacteristics}

Our numerical solutions may be characterized by several quantities. One of the most important of such quantities is the total mass $\mathcal M$. We compute $\mathcal M$ using the Komar expression \cite{PhysRev.113.934}, which for the axisymmetric spacetimes considered here takes the form
\begin{equation}
\label{eq:TotalMassEnergy}
\mathcal M
= 2 \pi \int_{\mathbb R^2} w(\rho,z) \rho\, \drho \dz
\end{equation}
where
\begin{equation}
w := e^{2\mu - 2\nu}\rho B T_{00} + \rho B (T_{\rho \rho} + T_{zz}) +  \frac{e^{2\mu + 2\nu}}{\rho B} T_{\varphi \varphi} - e^{2\mu - 2\nu} \rho B \omega^2 T_{\varphi \varphi}.
\end{equation}
We use the same letter $w$ for the integrand here as for the density in the Vlasov--Poisson case above. It should be clear from the context below which quantity is indicated. Note that for stationary asymptotically flat spacetimes the Komar mass is equivalent to the ADM mass \cite{Beig:1978vr,ChoquetBruhat:2008te}.
The mass plays an essential role in our iteration scheme below.  At each step of the iteration the ansatz function is renormalized such that the total mass $\mathcal M$ is unity. The Komar expression for the total mass of the system is derived from the time-symmetry of the spacetime. One also obtains a Komar integral expression based on the axial symmetry, namely the total angular momentum
\begin{equation}
\label{eq:TotalAngularMomentum}
\mathcal J = - 2\pi \int_{\mathbb R^2}   e^{2\mu - 2\nu}\rho B  \left( T_{0\varphi} + \omega T_{\varphi \varphi}   \right)\drho \dz.
\end{equation}

The above properties of a solution depend only on the symmetries of the spacetime and are independent of the matter model. In relativistic kinetic theory, there is also a divergence-free 4-vector called the particle current density
\begin{equation}
N^j(\rho,z) := \int_{\mathbb R^3} f_0(\rho,z,v) p^j  \frac{\mathrm{d}^3 v}{\sqrt{1 + |v|^2}},
\end{equation}
for $j=0,1,2,3$, where $p^j$ is computed from \Eqref{eq:VBasis}. We identify the zero component of the particle current density with the rest mass density, which for our axisymmetric solutions can be written
\begin{equation}
\label{eq:RestMassDensity}
N^0
= \int_{\mathbb R^3} f_0(\rho,z,v) e^{-\nu} \mathrm{d}^3 v
= \frac{2 \pi e^{-2\nu} }{B}\int_{e^\nu}^\infty \int_{-\overline s}^{\overline s} K\Phi(E, \rho s) (E - \omega \rho s) \ds \dE.
\end{equation}
The following quantity is then interpreted as the rest mass of the system,
\begin{equation}
\label{eq:RestMass}
 \mathcal M_0 = 2\pi \int_{\mathbb R^2} \rho B e^{2\mu} N^0 \drho \dz.
\end{equation}
Under time evolution both the total mass $\mathcal M$ and the rest mass $\mathcal M_0$ are conserved.

One of the most important issues concerning the time evolution of stationary solutions is the stability. In spherical symmetry there is numerical support \cite{Andreasson:2006dza} that the stability properties of static solutions is related to the normalized binding energy $E_b$ and the central redshift $Z_c$. These quantities are defined by the following expressions
\begin{align}
\label{eq:bindingenergy}
E_b =& 1 - \mathcal M / \mathcal M_0, \\
\label{eq:CentralRedShift}
Z_c =& \left( -g_{00}|_{\rho = z=0} \right)^{-1/2} - 1 = \left(\frac{1}{e^\nu \sqrt{1 - (\omega B \rho e^{-2\nu})^2}} - 1 \right)|_{\rho = z=0}.
\end{align}
Below, we record these quantities for the solutions which we compute in anticipation of future dynamical studies.

Another important measure of our solutions is the radius of support of the matter distribution. In spherical symmetry the ratio $2\mathcal M/R_0$, where $R_0$ is the radius of support in areal coordinates, is a measure of how relativistic a solution is.
It has been proved that for spherically symmetric regular bodies this quantity is bounded from above by $8/9$ \cite{Buchdahl:1959be,Andreasson:2008fu}, and it has also been proved that for the spherically symmetric Einstein--Vlasov system this bound is sharp \cite{Andreasson:2007kv}.

If we express the metric \Eqref{eq:Metric} in spherical coordinates, the radial coordinate $r := \sqrt{\rho^2 + z^2}$ is the isotropic radius. In spherical symmetry this can be related to the areal radial coordinate $R$ through
\begin{equation}
\label{eq:RCoord}
R = r ( 1+ \mathcal M/(2 r))^2.
\end{equation}
In this paper we use the coordinate $R$ defined by the above expression even in absense of spherical symmetry.
We denote the support of the matter by $R_0$, and in the isotropic radial coordinate by $r_0$.
For a spherically symmetric solution, the radius of support can be determined from the cutoff energy $E_0$ by matching the solution to a Schwarzschild exterior. The expression in terms of both the areal and isotropic coordinates is
\begin{equation}
\label{eq:E0andR0}
E_0 = \sqrt{1 - 2\mathcal M /R_0} = (1 - \mathcal M /(2 r_0))/ (1 + \mathcal M /(2 r_0)).
\end{equation}
Shapiro and Teukolsky use the quantity $R_0$ defined by \Eqref{eq:E0andR0} as a measure of how relativistic a solution is \cite{Shapiro:1993hi,Shapiro:1993gb}.

\section{Numerical Method}
\label{sec:NumericalMethods}
The numerical method used to solve the Vlasov--Poisson and
Einstein--Vlasov systems is a direct finite element discretization of
\Eqsref{eq:SemilinearPoisson}--\eqref{eq:VPDensityIntegral} and
\Eqsref{eq:EinsteEqNU}--\eqref{eq:EinsteEqWW}, respectively, in
combination with numerical integration of the matter terms. The
resulting system of nonlinear discrete equations is then solved using
a particular fixed-point iteration. We describe the finite element
discretization and fixed-point iteration in some detail below.

\subsection{Finite Element Method}
Let us briefly recall the finite element method (FEM) for solving a
boundary-value PDE problem \cite{Brenner:2008hf}. The idea is to formulate the boundary
value problem as a variational problem in a Sobolev space $V$ where
the solutions satisfy a corresponding weak-form of the equations.
Once a variational form has been obtained, one constructs a discrete
approximating subspace $V_h \subset V$ of the Sobolev space $V$ by
discretizing the solution domain and constructing a discrete
(finite-dimensional) function space on the resulting finite element
mesh, typically as a space of piecewise polynomial functions. One then
obtains a discrete system of equations by seeking the solution to the
variational problem on the discrete subspace $V_h$. If the original
PDE is linear, one obtains a linear system that can be solved using
either an iterative or direct solver, while if the original PDE is
nonlinear one obtains a nonlinear system that can be solved using an
iterative method such as direct fixed-point iteration or a Newton-type
method.

Before describing our finite element method, we note that the numerical solution domain is truncated to be the half-disk of radius $r_b$ in the meridional plane defined by
\[ D_{r_b} := \{ (\rho, z) : 0\le \rho^2 + z^2 \le r_b^2 \}. \]
We denote this by $D$ below when the specific radius is not relevant
for the discussion. Let
$\mathcal I_{r_b} := \left\{ (\rho, z) \in D_{r_b} : \rho^2 + z^2 =
  r_b^2 \right\} $
be the boundary of our solution domain which is to approximate spatial infinity, and
$\mathcal A = \left\{ (\rho, z) \in D_{r_b} : \rho = 0 \right\} $ be
the axis. As a consequence, the asymptotics for the gravitational potentials become boundary conditions strongly imposed at finite radius. Corresponding to \Eqref{eq:VlasovPoissonBCInfinity} for the Vlasov-Poisson system we have
\begin{equation}
\label{eq:NumericalBoundaryVP}
U_0 = - M/r_b \text{ on } \mathcal I_{r_b}
\end{equation}
and to \Eqref{eq:EinstBCInfinity} for the Einstein-Vlasov system we use
\begin{equation}
\label{eq:NumericalBoundaryEV}
\nu = - \mathcal M / r_b, \quad
B = 1, \quad
\mu = \mathcal M / r_b, \quad
\omega  = 0
\text{ on } \mathcal I_{r_b}.
\end{equation}
The approximation $\omega = 0$ on $\mathcal I_{r_b}$ is discussed further in \Sectionref{sec:ConsistencyChecks} below. 

The remainder of this section concerns the finite element method on the domain D. Let
$\langle u, v \rangle : = \int_D u v \drho \dz $ denote the
$L^2$-inner product on the solution domain $D$ and let
$\langle\langle u, v \rangle\rangle : = \int_D u v \rho \drho \dz $
denote the weighted inner product on $D$ reflecting the axial symmetry
of the problem. Further, we let $|||v||| = \sqrt{\langle\langle v, v
  \rangle\rangle}$ denote the corresponding norm, and we introduce
the weighted Sobolev space $W^1_{\rho}(D)$ defined by
\begin{equation}
  W^1_{\rho}(D)
  = \{ v : D \rightarrow \R : |||v|||^2 + |||\nabla v|||^2 < \infty \}.
\end{equation}

The weak form of the equations is formally obtained by multiplying the equations by test functions, integrating over the solution domain, and  transferring derivatives onto the test functions in the principle terms via integration by parts. In the Vlasov--Poisson case, the weak form of \Eqref{eq:SemilinearPoisson} reads
\begin{align}
\label{eq:UWeak}
\langle\nabla U_0, \rho \nabla v_0 \rangle  = -4 \pi \langle w_0(U_0), \rho v_0 \rangle,
\end{align}
with $w_0(U_0)$ given by \Eqref{eq:VPDensityIntegral}, and where $v_0$ is a test function in a space defined below.
For the Einstein--Vlasov system, the weak formulation of \Eqsref{eq:EinsteEqNU}-\eqref{eq:EinsteEqWW} is
\begin{align}
\label{eq:NUWeak}
\langle\nabla \nu, \rho \nabla v_1 \rangle
	&= 		-4\pi \langle\Phi_{00} + \Phi_{11}, \rho v_1\rangle \\ \nonumber
	& \quad	-4\pi \langle\left( 1 + (\rho B)^2 e^{-4 \nu} \omega^2 \right)\Phi_{33}, \rho v_1\rangle
			- 8\pi \langle e^{-4 \nu} \omega \Phi_{03}, \rho v_1\rangle	\\ \nonumber
	& \quad	+ \langle\frac 1B \nabla B \cdot \nabla \nu, \rho v_1\rangle
			- \langle \frac 12 e^{-4\nu} (\rho B)^2 \nabla \omega \cdot \nabla \omega , \rho v_1\rangle, \\
\langle \nabla B, \rho \nabla v_2\rangle - \langle B_{\rho}, v_2\rangle
	&= -8\pi \langle B \Phi_{11}, \rho v_2 \rangle, \\
\label{eq:MUWeak}
\langle \nabla \mu, \rho \nabla v_3\rangle + \langle \mu_{\rho}, v_3\rangle
	&= 		4\pi \langle \Phi_{00} + \Phi_{11} , \rho v_3\rangle, \\ \nonumber
	& \quad	+ 4\pi \langle \left((\rho B)^2 e^{-4 \nu} \omega^2 - 1 \right)\Phi_{33} + 2 e^{-4 \nu} \omega \Phi_{03} , \rho v_3\rangle, \\ \nonumber
	& \quad	- \langle \frac 1B \nabla B \cdot \nabla \nu, \rho v_3 \rangle
	      		- \langle \nu_{\rho}, v_3\rangle 	\\ \nonumber
	& \quad	+ \langle |\nabla \nu|^2, \rho v_3\rangle
			- \frac 14  \langle e^{-4\nu} (\rho B)^2 \nabla \omega \cdot \nabla \omega , \rho v_3\rangle, \\
\label{eq:WWWeak}
\langle \nabla \omega, \rho \nabla v_4\rangle - 2 \langle \omega_{\rho}, v_4\rangle
	&= - \langle \frac{16 \pi}{(\rho B)^2} \left( \Phi_{03} + (\rho B)^2 \omega \Phi_{33} \right), \rho v_4 \rangle \\ \nonumber
	& \quad
		+ \langle \frac 3B  \nabla B \cdot \nabla \omega , \rho v_4\rangle
		- \langle 4 \nabla \nu \cdot \nabla \omega, \rho v_4 \rangle,
\end{align}
where $v_1, v_2, v_3$ and $v_4$ are test functions.

\begin{remark}
  Due to the imposed axisymmetry of the problem, the integration over
  the two-dimensional domain $D_{r_b}$ is carried out with respect to
  the measure $2\pi\rho\drho\dz$. As a result, boundary terms on the
  axis $\mathcal{A}$ are \emph{natural} to the variational problem and
  vanish, whereas the boundary conditions on
  $\mathcal{I}_{r_b}$ are imposed \emph{strongly} on the finite
  element function space.
\end{remark}

The variational formulations for the Vlasov--Poisson and Einstein--Vlasov systems read as follows.
\begin{definition}[Variational Vlasov--Poisson Problem]
Find $U_0$ in the space
\begin{equation}
V = \left\{ v \in W^1_{\rho}(D_{r_b}): v  \text{ satisfies} \text{ \Eqref{eq:NumericalBoundaryVP}}
\right\}
\end{equation}
such that the variational problem \Eqref{eq:UWeak} is satisfied for
all test functions $v_0$ in the space
\begin{equation}
\hat{V} = \left\{ v \in W^1_{\rho}(D): v = 0 \text{ on } \mathcal I_{r_b}
\right\}.
\end{equation}
\end{definition}

\begin{definition}[Variational Einstein--Vlasov Problem]
Find $(\nu, B, \mu, \omega)$ in the space
\begin{equation}
V = \left\{ v \in \left[W^1_{\rho}(D)\right]^4: v \text{ satisfies}
\text{ \Eqref{eq:NumericalBoundaryEV} and \Eqref{eq:EinstBCAxis}}
\right\}
\end{equation}
such that the variational problem
\Eqsref{eq:NUWeak}--\eqref{eq:WWWeak}
is satisfied for all test functions $v = (v_1, v_2, v_3, v_4)$ in the
space
\begin{equation}
\hat{V} = \left\{ v \in \left[W^1_{\rho}(D)\right]^4:
 v_1 = v_2 = v_3 = v_4 = 0 \text{ on } \mathcal I_{r_b} \text{ and }
 v_3 = 0 \text{ on } \mathcal A
\right\}.
\end{equation}
\end{definition}

These variational problems are discretized using a finite dimensional subspace $V_h \subset V$, which is spanned by
piecewise polynomial functions over an unstructured triangular mesh on D. The mesh is taken to be large compared to the support of the matter -- typically taken the radius to be $r_b = 50$ for solutions with zero net angular momentum and $r_b = 100$ for solutions with non-zero net angular momentum -- and is refined in the region of matter support. Although we are exploring fully adaptive mesh schemes presently, the meshes used here are generated a~priori to match solution characteristics.

\subsection{Numerical Integration}

The matter terms Eq.~\eqref{eq:VPDensityIntegral} and
\Eqsref{eq:Phi00_expression}--\eqref{eq:Phi03_expression} appearing in
the discretized systems are evaluated using numerical integration; at
each nodal point of the finite element mesh, the matter terms are
integrated numerically using a basic second-order accurate midpoint
scheme in both integration variables. The number of integration steps
in each dimension is tuned for numerical accuracy. We have typically
used a value of $32$ integration steps for the simulations in the
present paper.

\subsection{Fixed-Point Iteration}
The task of the numerics is to find a self-consistent solution to
\Eqsref{eq:SemilinearPoisson}--\eqref{eq:VPDensityIntegral} in the
Vlasov--Poisson case, and
\Eqsref{eq:EinsteEqNU}--\eqref{eq:EinsteEqWW} with the matter terms
\Eqsref{eq:Phi00_expression}--\eqref{eq:Phi03_expression} in the
Einstein--Vlasov case. This is achieved with an iteration procedure in
which the total mass is fixed at each iteration. We start by
prescribing an initial guess for the potential $U_0$, or the metric
fields $\nu, \mu, B$, and $\omega$. These initial potentials are then
used to compute the matter terms for a given ansatz function with unit
normalization constant $K$ (cf.~\Eqref{eq:AnsatzForm}). The constant $K$ is fixed by the
constraint that the total mass be the prescribed value. At this
stage we have a coupled set of linear elliptic equations for the
gravitational potentials ($U_0$ in the Vlasov--Poisson case, and
$\nu, \mu, B$, and $\omega$ in the Einstein--Vlasov case), which we
solve using the finite element method described above implemented in
FEniCS~\cite{Logg:2012jw,Logg:2010kt}. Once the linear system of
equations has been solved, the matter terms are evaluated, the constant $K$ is fixed
once again, and the procedure is iterated to convergence. The tolerance for convergence is set to $10^{-4}$.

\begin{remark}
Because the normalization constant $K$ in the ansatz is changed at each step of the iteration, the exact problem we solve is not determined until the end of the iteration. Although the functional form of the ansatz $\Phi(E,L_z)$ is specified, it only becomes apparent in the iteration which member of this family has the prescribed mass.
\end{remark}

The initial guess for the above iteration may be either a rough
estimate, or a previously computed solution. Many solutions that our
code obtains are robust against variations in the initial
guess. However, for the code to converge to more extreme solutions it
may be required that the initial guess is sufficiently close. In such
cases, the extreme solution is approached by first solving for a
series of intermediate solutions. In certain cases a damped
fixed-point method is also used to obtain convergence. Our code
presently implements a linear damping scheme of the form
\begin{displaymath}
  X_n = (1 - \theta) X_{n-1} + \theta \mathcal Y(X_{n-1}),
\end{displaymath}
where $X$ denotes the vector of degrees of freedom for the metric
fields, $\mathcal Y$ denotes the fixed-point iteration, including solution of
the linear system obtained by the finite element discretization, the
numerical integration and the rescaling of the constant $K$, and where
$\theta \in (0, 1]$ is a parameter. By default, we set $\theta = 1$
and reduce the value of $\theta$ in cases when the fixed-point
iteration fails to converge. In many cases, this extends the model
regime for which solutions can be obtained.

\begin{remark}
\label{rem:PreferStable}
In our numerical experiments in the axial symmetric case, as well as in a similar algorithm implemented in spherical symmetry, we observe that the iteration appears to converge to dynamically stable solutions. This is particularly evident in the spherically symmetric case, where the stability of equilibrium solutions has been numerically investigated \cite{Andreasson:2006dza}.
We also note that a similar relation has been observed in the work of Andr\'easson and Rein in the flat case \cite{Andreasson:2014gb}, where a similar algorithm is used.
\end{remark}

\subsection{Consistency Checks}
\label{sec:ConsistencyChecks}
In the discussion above we focus on an implementation which uses \Eqref{eq:EinsteEqMU} for the field $\mu$, rather than the equation for $\xi$ (cf.~\Eqref{eq:XIequation}). Indeed, because the form of \Eqref{eq:EinsteEqMU} matches those for the other metric fields (as well as Newtonian potential) we implement this equation in most of our simulations. We have verified however that the same results are obtained if one instead integrates \Eqref{eq:XIequation}.

Unlike the total mass of the solution, the total angular momentum is
not prescribed, but rather computed via
\Eqref{eq:TotalAngularMomentum} once the iteration has
converged. During the iteration the boundary condition for the field
$\omega$ is set to zero (cf. \Eqref{eq:NumericalBoundaryEV}), introducing an error of $2\mathcal J/r_b^3$
in the numerical boundary condition. For the rotating solutions we
compute $\mathcal J \sim 1.0$ so this error is typically of the order
$\sim 10^{-6}$ if we take $r_b = 100$. What effect does this have on
the solution? To estimate this error we consider the case of a
rotating toroidal solutions with high angular momentum. On the one
hand we compute the solution with vanishing boundary conditions for
the $\omega$ field, and on the other we iterate the solution procedure
updating the angular momentum in the boundary condition based on the
value computed in the previous iterate. We do this until the computed
angular momentum converges to within a tolerance of $10^{-4}$. The
error in the solution is estimated via the normalized $L^2$ norm of
the differences of the metric fields. For the $\omega$ field this
error is $2.1 \cdot 10^{-2}$, (i.e. at the one-percent level), while for the $\nu, \mu$ and $B$ fields it is order $10^{-5}$.
Since the solution in which the boundary value of the total angular momentum is iterated is much more computationally expensive, and since the two solutions (with and without the iterated boundary conditions) differ only at the one percent level, and do not exhibit any significant differences, we take $\mathcal J = 0$ on the boundary (cf.~\Eqref{eq:EinstBCInfinity}) in the remaining runs.

A further check on the numerics can be made by comparing the total mass and angular momentum computed with the integral expressions \Eqref{eq:TotalMassEnergy} and \Eqref{eq:TotalAngularMomentum} with values read off from the asymptotic behavior of the metric fields. More precisely we have \cite{Shapiro:1993hi}
\[
 \mathcal M_{\mathrm{inf}} =\lim_{r\to \infty} \frac 12 r (1-e^{2\nu}), \quad \mathcal J_{\mathrm{inf}} = \lim_{r\to \infty} \frac 12 r^3 \omega B^2 e^{-2\nu} .
\]
We verify that the value $\mathcal J_{\mathrm{inf}}$ computed from the expression above approaches the angular momentum computed with \Eqref{eq:TotalAngularMomentum} at some radius in between the radius of support and the domain boundary. The error is  a few percent for a domain of radius $r_b = 100$. Clearly, since the total mass is prescribed and used in the boundary condition, the value $\mathcal M_{\mathrm{inf}} $ approaches $\mathcal M$ near the boundary.

In our computations we take the total mass $\mathcal M = 1$, which is equivalent to scaling out of the equations the only dimensionful quantity. However, we also check that the solutions scale as expected when the prescribed mass is increased; that is, increasing $\mathcal M $, while keeping the particle mass fixed at $m=1$, and simultaneously changing the parameter $L_0$ to $\mathcal M L_0$, we find the radius of support and total angular momentum to increase according to $r_0 \rightarrow \mathcal M r_0$ and $\mathcal J \rightarrow \mathcal M^2 \mathcal J$.

A good test for our code is in comparing to solutions obtained in \cite{Shapiro:1993gb,Shapiro:1993hi}. By modifying our ansatz function to match that used by Shapiro and Teukolsky and appropriately scaling the solution characteristics we find agreement, for example, with the results of Table 2 in \cite{Shapiro:1993gb}.

\section{Results}
\label{sec:Results}
In numerical results below we use an ansatz $K\Phi(E,L_z)$ with a product structure
\begin{equation}
\label{eq:ProductAnsatz}
 \Phi(E, L_z) = \phi(E) \psi(L_z),
\end{equation}
with various choices of $\phi$ and $\psi$. If $\psi$ is an even function, there are an equal number particles rotating in each direction; the solution will then have zero net angular momentum. We also consider ansatz functions for which $\psi(L_z)$ vanishes for $L_z < 0$, thus forcing all particles to have angular momentum of the same sign. Solutions generated by such an ansatz have a net angular momentum. While all of the ansatzes considered in this paper have this product structure, except for the composite solutions considered in \Sectionref{sec:CompositeObjects}, which are the sum of such product ansatzes, one does not generally have to make this choice.

\subsection{Toroidal Solutions}
\label{sec:PolytropicAnsatzSolutions}
We begin by presenting solutions generated by a four-parameter family of ansatz functions, which generalize the well-known polytropic ansatz. These take the product form above with
\begin{equation}
\label{eq:PolytropicEAnsatz}
\phi(E) =
\begin{cases}
       (E_0 - E)^k, & E \leq E_0, \\
       0,  		     &	E > E_0, \\
   \end{cases}
\end{equation}
and
\begin{equation}
\label{eq:PolytropicLAnsatz}
\psi(L_z) =
\begin{cases}
       (|L_z| - L_0)^l, & |L_z| > L_0, \\
       0,  		     &	|L_z| \leq L_0. \\
   \end{cases}
\end{equation}

\begin{figure}[htb!]
    \centering
  \begin{subfigure}[b]{0.35\linewidth}
    \centering
    \includegraphics[width=0.95\linewidth]{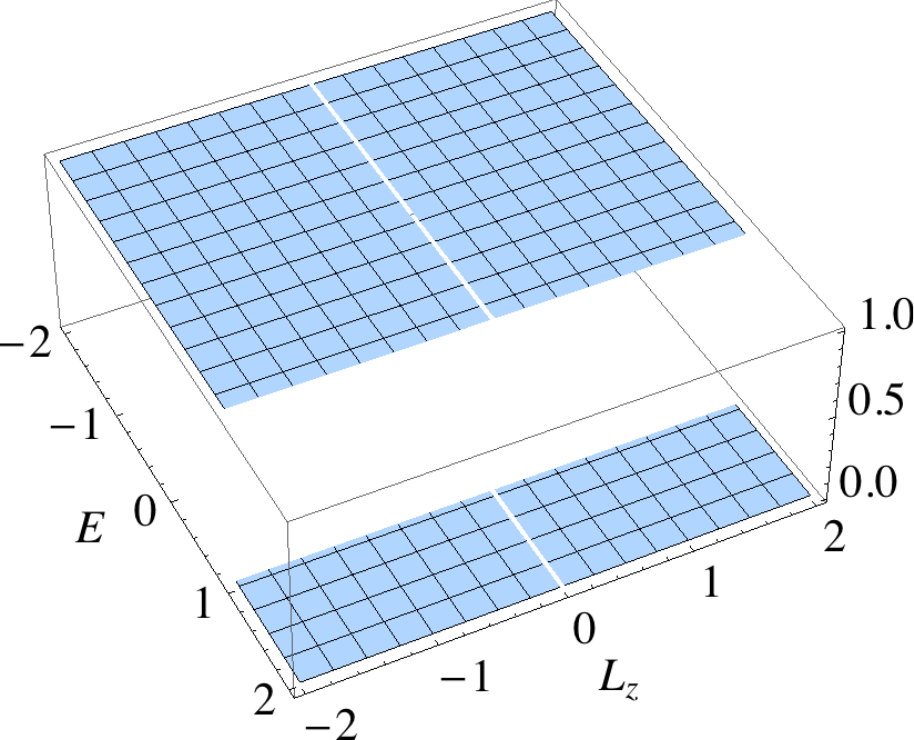}
    \caption{\small{Ansatz GP-A ($k=l=L_0=0$)}}
        \label{fig:evpolytrope:a}
  \end{subfigure}
  \begin{subfigure}[b]{0.35\linewidth}
    \centering
    \includegraphics[width=0.95\linewidth]{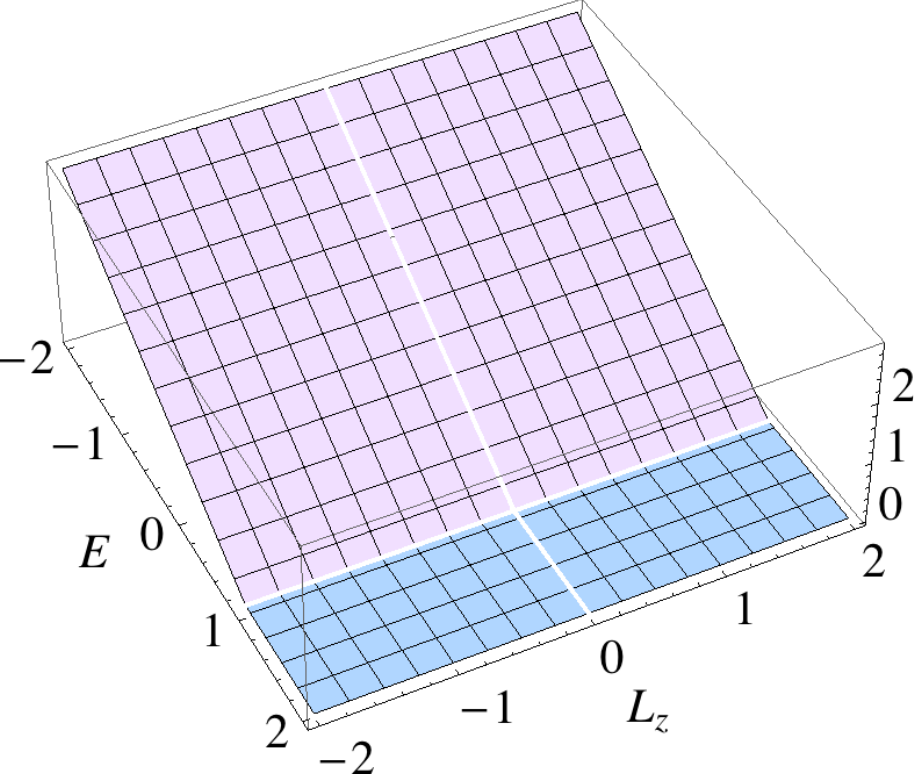}
    \caption{\small{Ansatz GP-B ($k=1, l=L_0=0$)}}
    \label{fig:evpolytrope:b}
  \end{subfigure}
  \begin{subfigure}[b]{0.35\linewidth}
    \centering
    \includegraphics[width=0.95\linewidth]{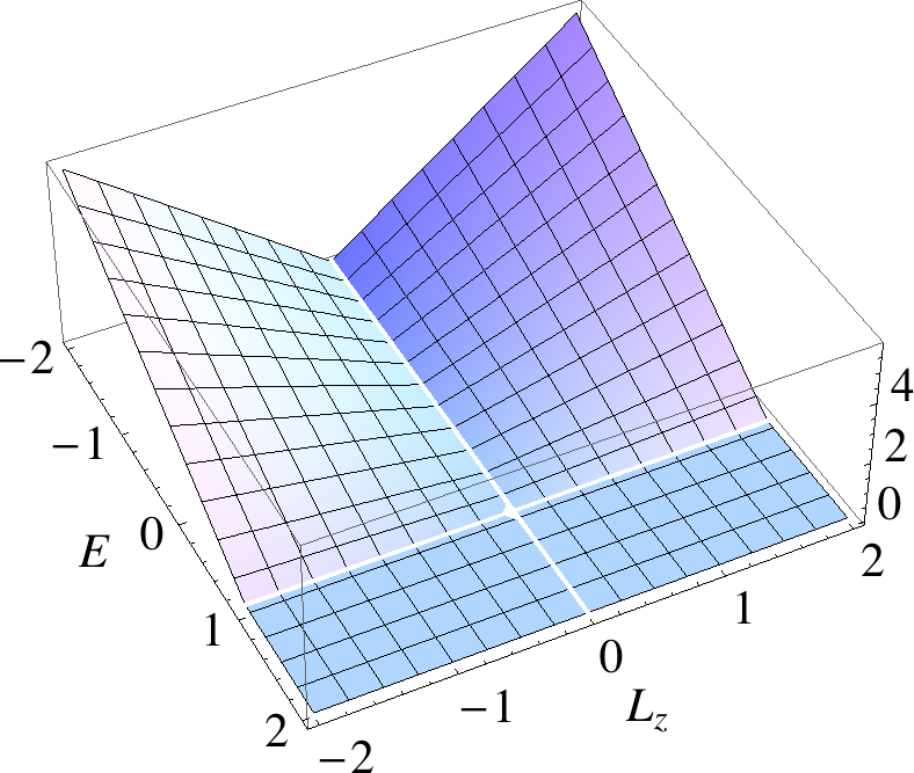}
    \caption{\small{Ansatz GP-C ($k= l =1, L_0=0$)}}
    \label{fig:evpolytrope:c}
  \end{subfigure}
  \begin{subfigure}[b]{0.35\linewidth}
    \centering
    \includegraphics[width=0.95\linewidth]{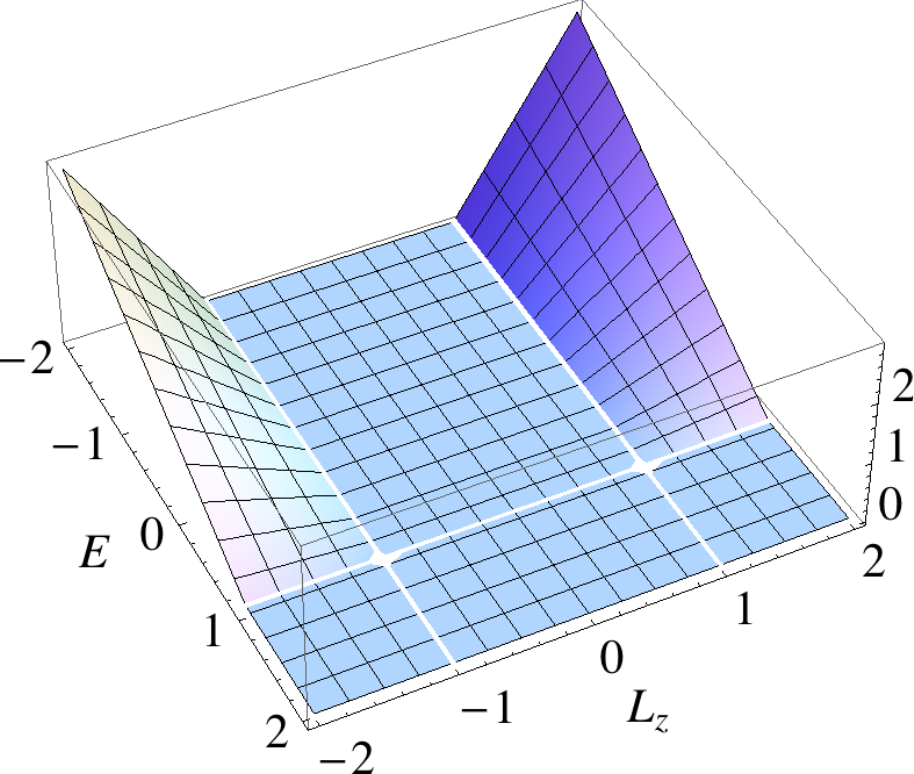}
    \caption{\small{Ansatz GP-D ($k=l=L_0=1$)}}
    \label{fig:evpolytrope:d}
  \end{subfigure}
\captionsetup{width=.7\linewidth}
  \caption{
  Ansatz functions for the generalized polytropic solutions. GP-A (Panel \ref{fig:evpolytrope:a}) and GP-B (Panel \ref{fig:evpolytrope:b}) are independent of $L_z$ and result in spherically symmetric solutions. GP-C (Panel \ref{fig:evpolytrope:c}) and GP-D (Panel \ref{fig:evpolytrope:d}) break the spherical symmetry and generate spatial densities which vanish at the axis or with vacuum at the axis respectively.
  }
  \label{fig:evpolytrope}
\end{figure}

The ansatz has four parameters, a cut-off energy $E_0$, exponents $k$ and $l$, and an angular momentum cut-off $L_0$.
The choice of $L_0 = 0$ corresponds to the familiar polytropic solutions (see for example \cite{binney2011galactic}), and motivates calling this ansatz \emph{generalized polytropic}. Energy-angular momentum phase-space plots of the ansatz functions in four illustrative cases are found in \Figref{fig:evpolytrope}. These four choices of the ansatz parameters allow us to demonstrate the basic dependence of the solutions on the four parameters, and also to verify that the code gives reasonable results in the well-studied spherically symmetric case.
We illustrate these in the Einstein--Vlasov model with zero net angular momentum, and note that the forms of the ansatz and corresponding spatial densities are the same in the Einstein--Vlasov model with net angular momentum and in the Vlasov--Poisson model.
In the following section we investigate the limits of our method in producing relativistic rotating toroidal configurations.

\begin{figure}[htb!]
\centering
  \centering
  \includegraphics[width=.7\linewidth]{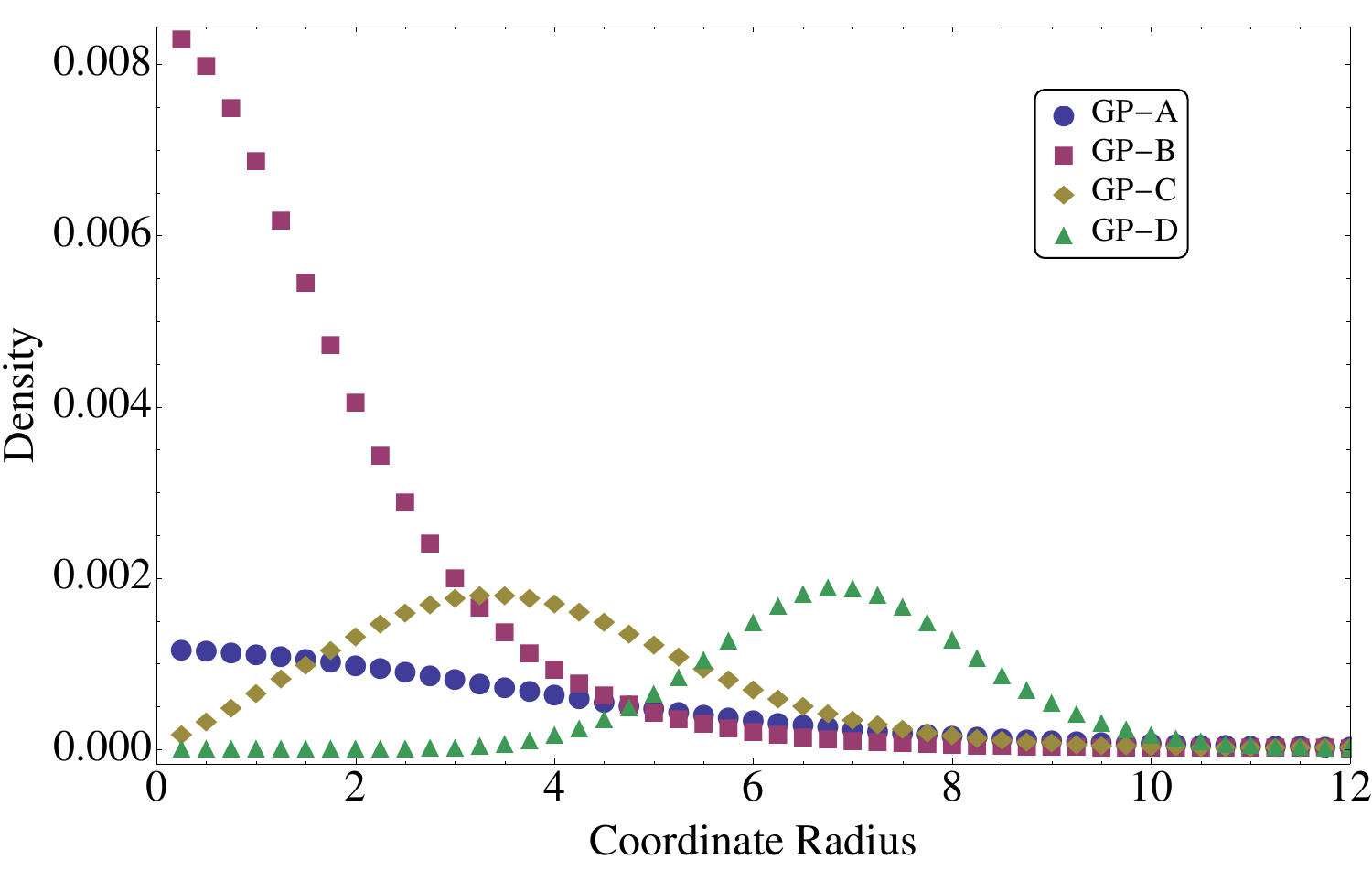}
  \captionsetup{width=.7\linewidth}
 \caption{Density profiles for the generalized polytropic solutions in a slice through $z=0$ versus coordinate radius $\rho$. The radius of support for the solutions listed in \Tableref{table:PolytropicSolutions} is given in terms of the coordinate $R$, cf.~\Eqref{eq:RCoord}.
 }
  \label{fig:PolytropicDensityTraces}
\end{figure}

Solution parameters and characteristics for the four cases demonstrated here are collected in \Tableref{table:PolytropicSolutions}. In \Figref{fig:PolytropicDensityTraces} we show the spatial density for a $z=0$ trace in the meridional plane.
A spherical distribution is obtained when all values of angular momentum are equally weighted by the ansatz function, which can be see in cases GP-A and GP-B (generalized polytrope A and B respectively). The solution GP-B is more centrally condensed than GP-A due to the fact that in the ansatz function GP-B, higher energy particles are relatively suppressed. However, the radius of support for these solutions is the same, and since these solutions are spherically symmetric, can be determined from \Eqref{eq:E0andR0}. For the solutions presented here $E_0 = 0.925$, and $\mathcal M = 1$, giving $R_0 =  13.853$. Indeed, this value is observed in the solutions (see  \Tableref{table:PolytropicSolutions}). We also note that the solutions exhibited here are not very relativistic in the sense that the value $2\mathcal M /R_0$ is far from the Buchdahl bound of $8/9$ \cite{Buchdahl:1959be,Andreasson:2008fu}.

\begin{table}[htp]
  \begin{center}
  \begin{threeparttable}
  \footnotesize
    \begin{tabular}{llcc|ccccccc}
      \toprule
     \textbf{Model}  & \multicolumn{3}{c}{\textbf{Parameters}} 	&  \multicolumn{7}{c}{\textbf{Solution Characteristics}}   \\
      Einstein--Vlasov 		    & $k$ & $l$ & $L_0$ 				   	& $w_p$ & $\rho_p$ & $K^{-1}$  & $E_b$ & $Z_c$ & $R_0$ & $2\mathcal{M}/R_0$ \\
            \midrule
      GP-A    & $0.0$ & $0.0$ & $0.0$ 						&  $0.0011$ & $0.0$ & $1108.10$ & $0.027$ & $0.235$ &  $13.86$  & $0.144$  \\
                 [1em]
      GP-B    & $1.0$ & $0.0$ & $0.0$ 					&  $0.0084$ & $0.0$ & $81.44$ & $0.037$ & $0.443$ & $13.85$ & $0.144$  \\
                 [1em]
      GP-C    & $1.0$ & $1.0$ & $0.0$ 					&  $0.0018$ & $3.33$ & $27.42$ & $0.032$ & $0.250$ & $14.12$ & $0.142$  \\
                 [1em]
      GP-D    & $1.0$ & $1.0$ & $1.0$ 					&  $0.0019$ & $6.84$ & $3.02$ & $0.027$ & $0.151$ & $14.43$ & $0.139$  \\
      \bottomrule
    \end{tabular}
  \caption{
A comparison of the generalized polytropic solutions. All solutions have $E_0 = 0.925$, and $\mathcal M = 1.0$. The peak density $w_p$ occurs at coordinate value $\rho_p$. $K$ is the coefficient appearing in the ansatz function. $E_b$ and $Z_c$ are the normalized binding energy and central redshift.
}
\label{table:PolytropicSolutions}
\end{threeparttable}
\end{center}
\end{table}

The spherical symmetry can be broken by introducing dependence on $L_z$, either by increasing $l$ or $L_0$.
In the former case, GP-C, the peak density is shifted away from the coordinate origin into a ring and the spatial density vanishes asymptotically at the center of the configuration. The latter case leads to a toroidal distribution with vacuum in the center; this is illustrated by GP-D, where the density vanishes at approximately $\rho \approx 2.25 $. We observe that the relationship \Eqref{eq:E0andR0} between the cut-off energy and the radius of support does not generally hold in axisymmetry.

\subsection{Thin Toroidal Solutions}
\label{sec:ThinTorusSolutions}
Since the rigorous existence of axisymmetric solutions to the Einstein--Vlasov system is known only for solutions which are close to spherically symmetric Newtonian solutions, it is of interest to investigate the limit of how relativistic solutions we are able to construct using the present numerical method. We investigate this limit using a rotating version of the generalized polytropic ansatz, taking $L_0$ non-zero; that is, the $\psi(L_z)$-part of the ansatz takes the form
\begin{equation}
\label{eq:PolytropicLAnsatzRotating}
\psi(L_z) =
\begin{cases}
       (L_z - L_0)^l, & L_z > L_0, \\
       0,  		     &	L_z \leq L_0. \\
   \end{cases}
\end{equation}
This differs from \Eqref{eq:PolytropicLAnsatz} in that only particles with positive angular momentum $L_z$ are allowed.
In our trials with a number of different ansatz functions, this ansatz -- including the presence of net rotation -- was the most successful in constructing relativistic solutions. In the present paper we construct a sequence of solutions from the product ansatz given by \Eqsref{eq:PolytropicEAnsatz}-\eqref{eq:PolytropicLAnsatzRotating}, with decreasing $E_0$ parameter. Plots of the density $w$ for the more extreme members of a particular sequence with parameters $L_0 = 0.8, k=l=0$ are shown in \Figref{fig:TorusDensity} below.

\begin{figure}[htb!]
\centering
  \centering
  \includegraphics[width=.7\linewidth]{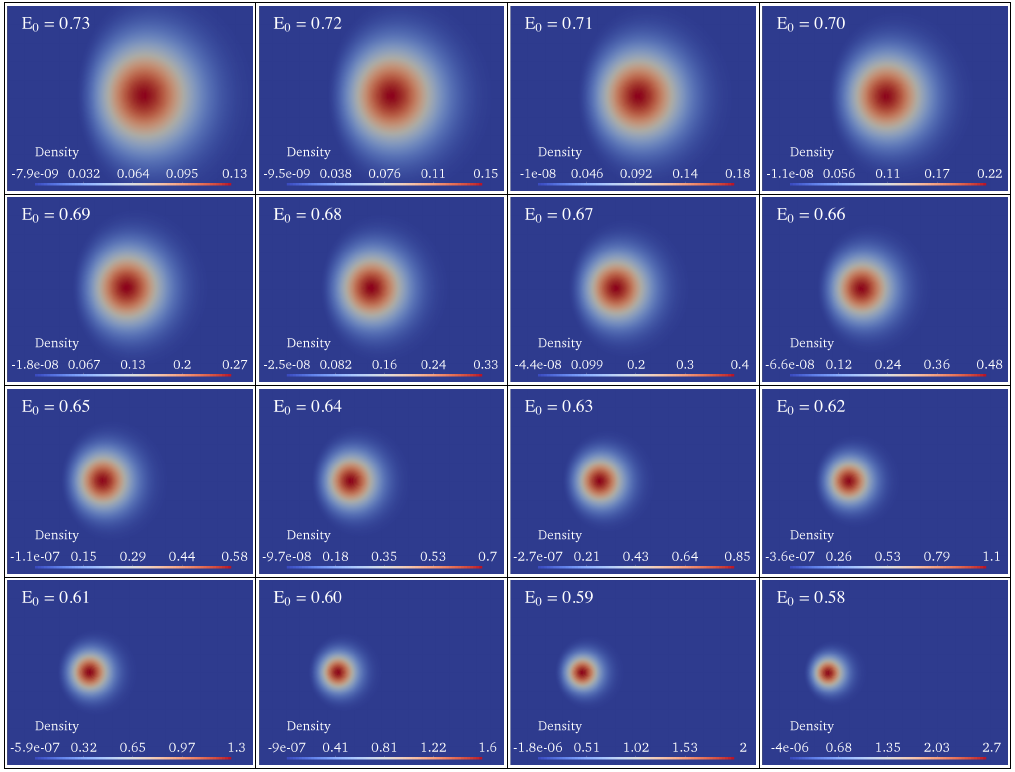}
  \captionsetup{width=.7\linewidth}
  \caption{Density plots for the limiting portion of a sequence of relativistic rotating toroidal solutions. The ansatz parameters in this case are $L_0 = 0.8$ and $k=l=0$.}
\label{fig:TorusDensity}
\end{figure}

\begin{figure}[htb!]
    \centering
  \begin{subfigure}[b]{0.35\linewidth}
    \centering
    \includegraphics[width=0.95\linewidth]{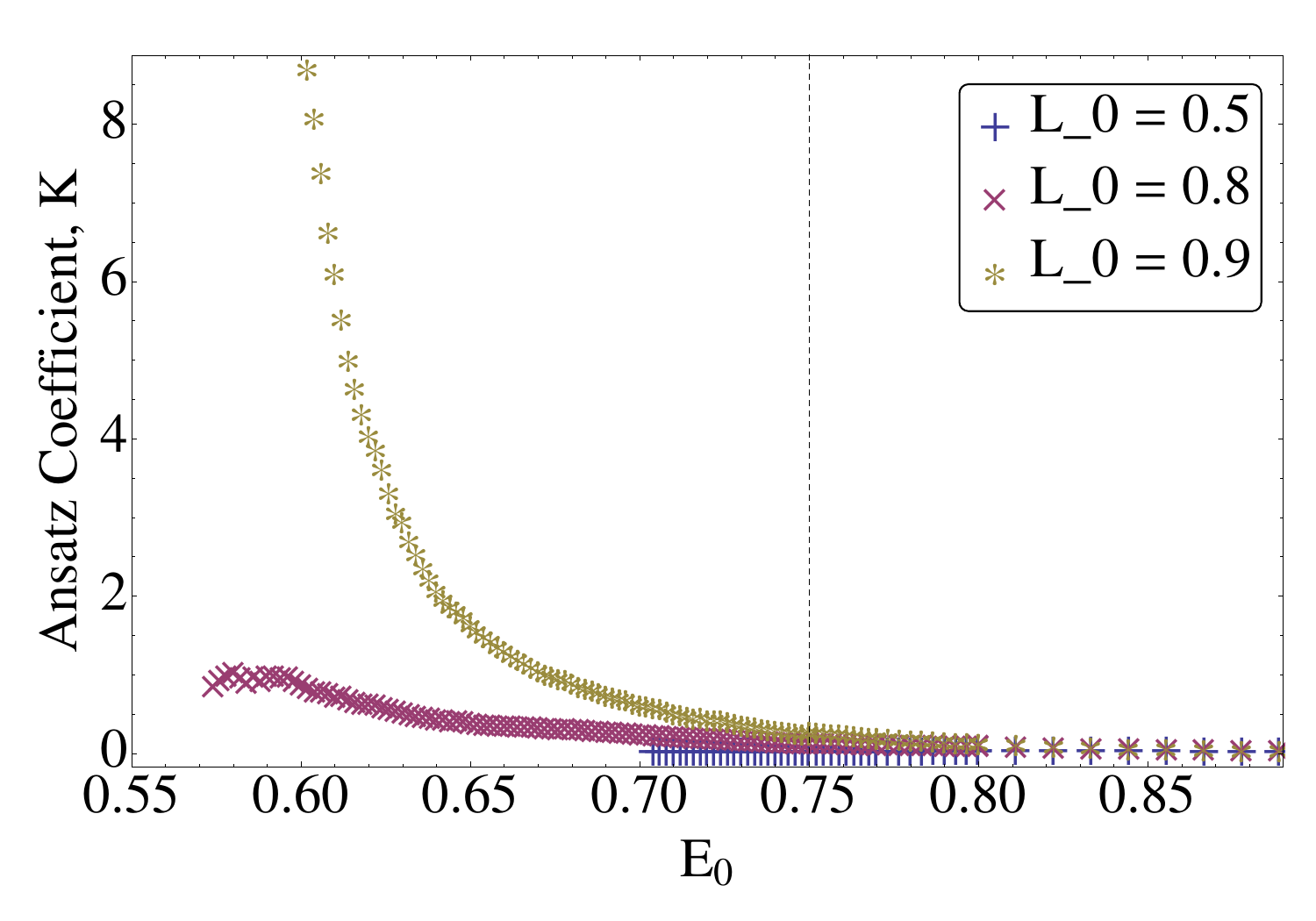}
    \caption{}
    \label{fig:KvsE0_subplot}
  \end{subfigure}
  \begin{subfigure}[b]{0.35\linewidth}
    \centering
    \includegraphics[width=0.95\linewidth]{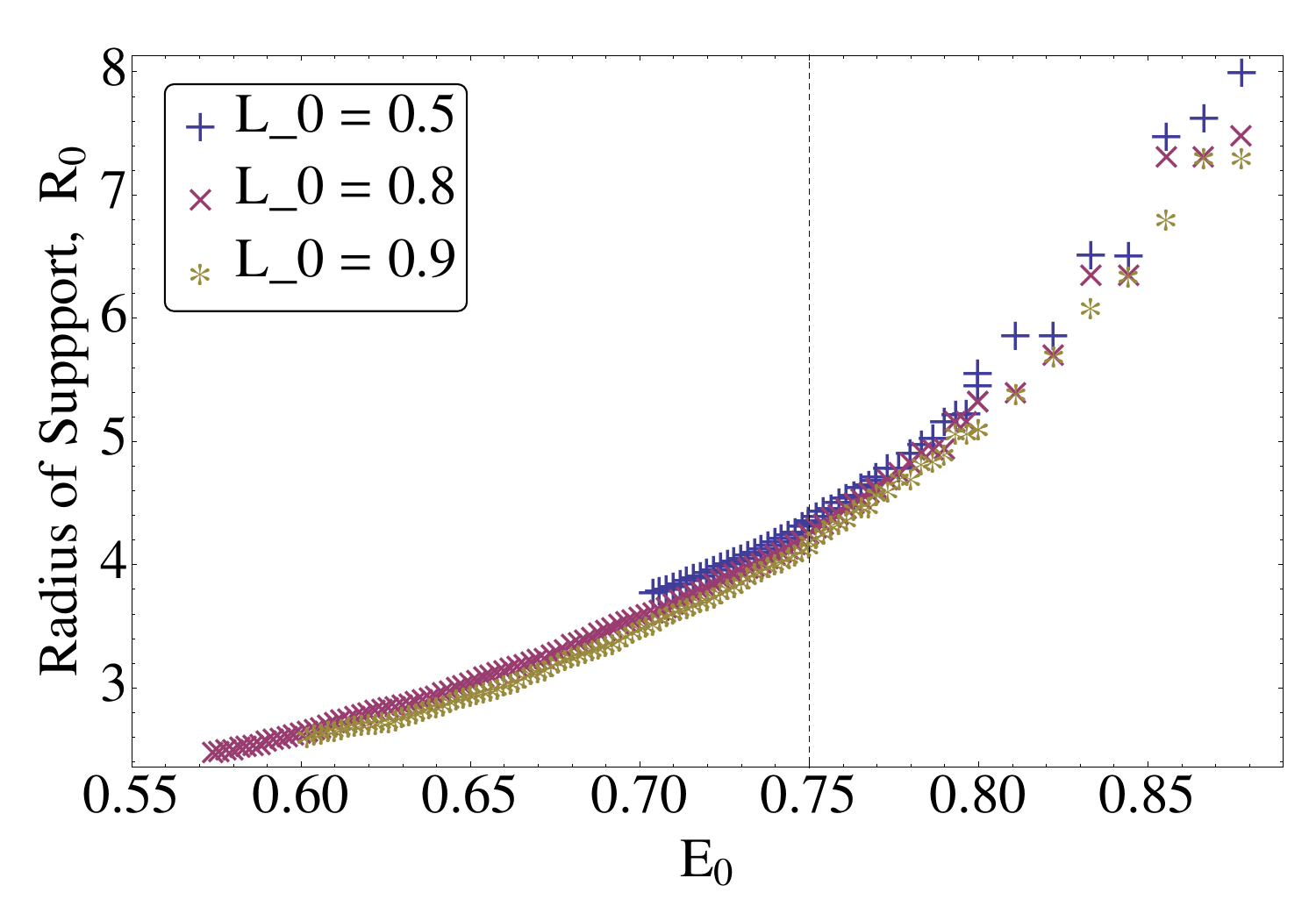}
        \caption{}
            \label{fig:R0vsE0_subplot}
  \end{subfigure}
  \begin{subfigure}[b]{0.35\linewidth}
    \centering
    \includegraphics[width=0.95\linewidth]{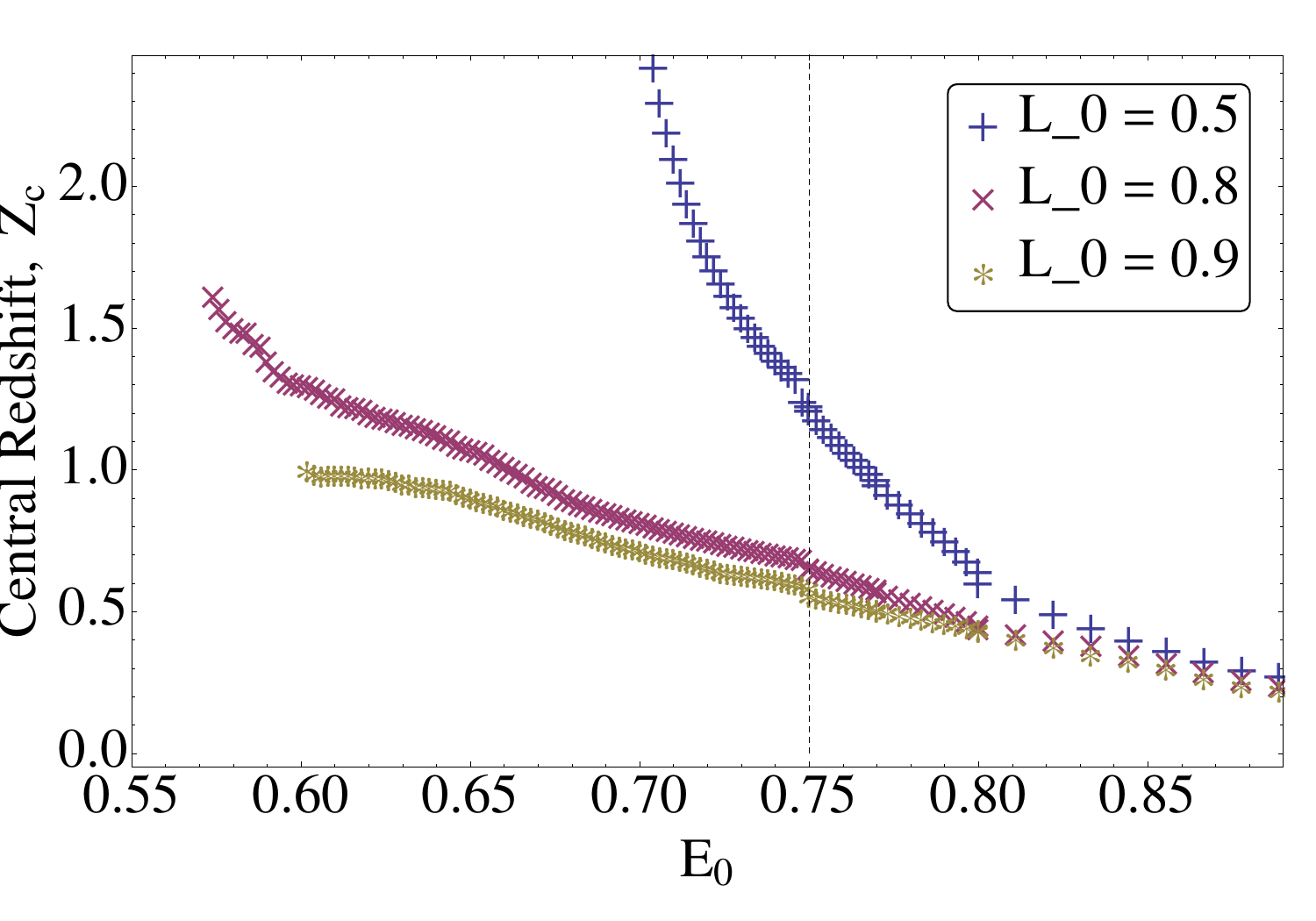}
        \caption{}
            \label{fig:ZcvsE0_subplot}
  \end{subfigure}
  \begin{subfigure}[b]{0.35\linewidth}
    \centering
    \includegraphics[width=0.95\linewidth]{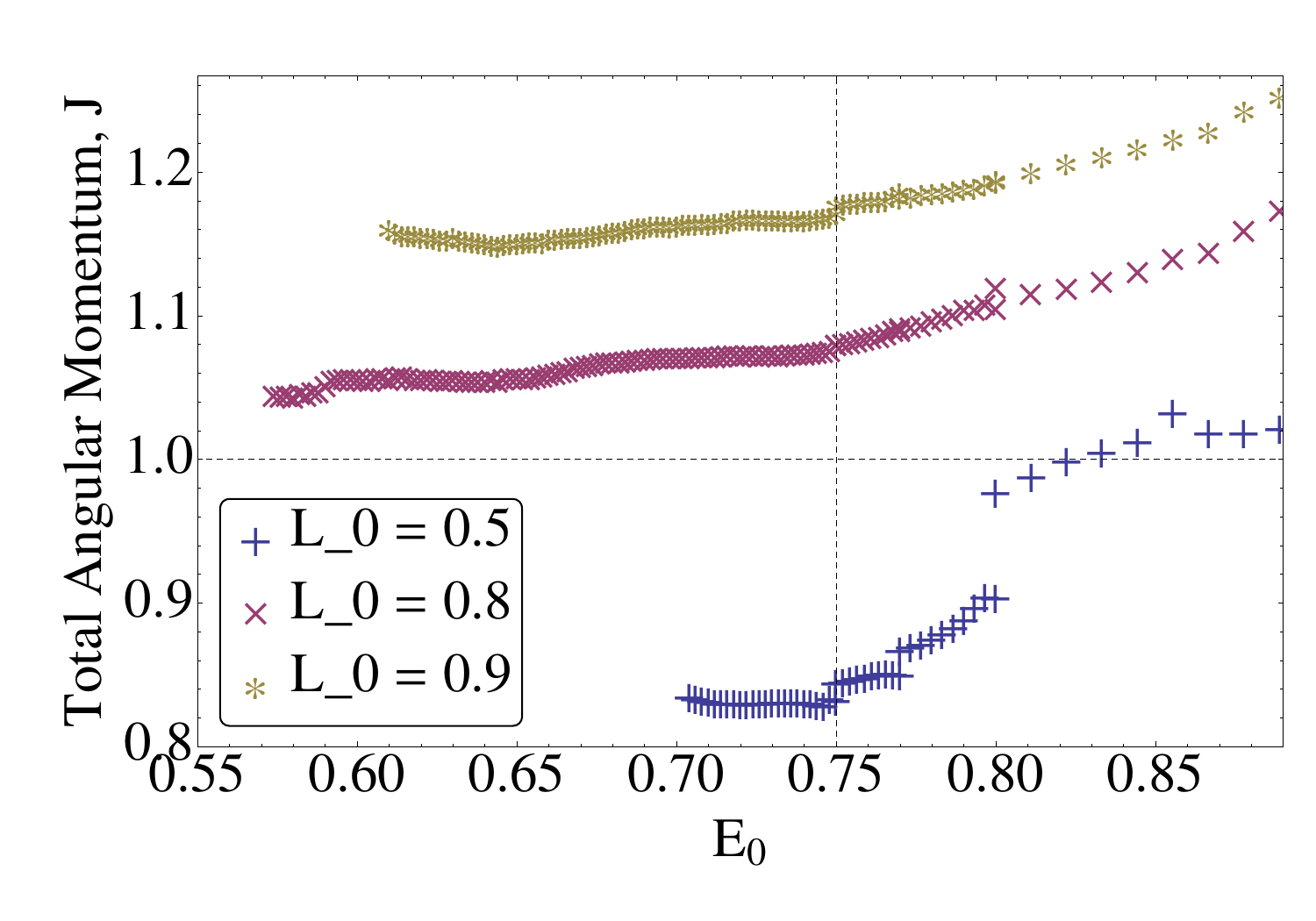}
            \caption{}
                \label{fig:JvsE0_subplot}
  \end{subfigure}
  \captionsetup{width=0.7\linewidth}
  \caption{Solution characteristics versus $E_0$ parameter for $L_0=(0.5,0.8,0.9)$ rotating torus solution sequences in blue ``$+$", red ``$\times$", and yellow ``$\Asterisk$" respectively. The upper left panel shows the ansatz coefficient $K$, the lower left shows the central redshift $Z_c$, the upper right shows the radius of support $R_0$ cf.~\Eqref{eq:RCoord}, and the lower left displays the total angular momentum $\mathcal J$ computed via \Eqref{eq:TotalAngularMomentum}. Solutions with $E_0 > 0.75$ are not fully converged -- see text.
  }
  \label{fig:EVR_rel_torus_solchars}
\end{figure}

Our most relativistic solution has parameters $E_0 = 0.58$, and $L_0 = 0.8, k=l = 0$. The actual radius of support of this solution is found to be $R_0 = 2.43$ (cf.~\Figref{fig:EVR_rel_torus_solchars} upper right panel), which is less than the value $R_0 = 2.96$ which one would obtain via \Eqref{eq:E0andR0} in spherical symmetry.
Interestingly, we note that the non-spherically symmetric solutions in \Tableref{table:PolytropicSolutions} have a radius of support which is larger than that of the spherically symmetric ones. This property is likely caused by the high angular momentum of this rotating solution. For spherically symmetric bodies, the compactness $2 \mathcal M /R_0$ gives a good measure of how relativistic the solutions are. While the meaning of this ratio is not as clear in axisymmetry, we note that the rotating toroidal solution which we construct here has $2\mathcal M /R_0 = 0.82$, which is close to the limiting value in spherical symmetry of $8/9$.

\begin{figure}[htb!]
\centering
  \centering
  \includegraphics[width=.7\linewidth]{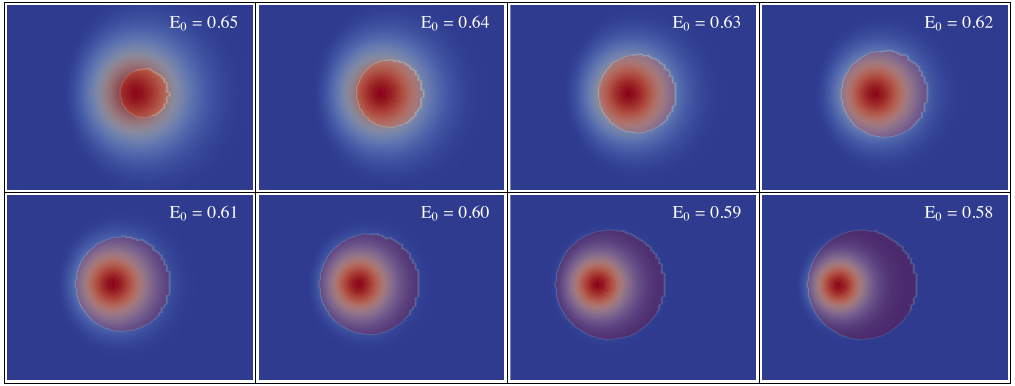}
\captionsetup{width=.7\linewidth}
  \caption{The development of the ergoregion in the $L_0 = 0.8$ solution family.}
\label{fig:ErgoRegionGrid}
\end{figure}

Perhaps a stronger indication that this sequence of solutions is relativistic is the presence of ergoregions. An ergoregion is a region of a rotating spacetime in which the Killing vector field $\partial_t$ is spacelike, forcing all causal observers to be dragged along in the direction of rotation. These regions are familiar from the Kerr-family black holes. The solutions presented here, are however, perfectly regular. As we show in \Figref{fig:ErgoRegionGrid}, the ergoregion begins to form within the matter at around $E_0 = 0.65$ and grows, in subsequent solutions, to eventually contain the entire support of the matter. These results lend strong support to the existence of solutions to the Einstein--Vlasov system with ergoregions, a question which was left open in the only rigorous existence proof for axisymmetric solutions \cite{Andreasson:ch}. Regular (non-black hole) solutions which contain ergoregions have been numerically constructed before in the case of uniformly rotating fluid models \cite{Fischer:2005bw,Ansorg:2003dk,Schbel:2003em}.

While the sequence of solutions discussed above leads to the solution with the lowest $E_0$ parameter, we also investigate sequences with greater and lesser angular momentum, which we control by adjusting the parameter $L_0$. A comparison of different solution characteristics for families of solutions with $L_0 = (0.5, 0.8, 0.9)$ is presented in \Figref{fig:EVR_rel_torus_solchars}. As illustrated in \Figref{fig:JvsE0_subplot}, the sequences with total angular momentum  $\mathcal J$ greater than the mass $\mathcal M$ squared can be extended to much lower $E_0$ parameter values. In view of \Remarkref{rem:PreferStable}  we conjecture that this is because these solutions are in the super-extremal regime where no Kerr black holes exist, and therefore may be stable, while the solution sequences with $\mathcal J < \mathcal M^2$ likely collapse to a Kerr black hole as they become sufficiently relativistic. In \cite{Abrahams:1994kl} it is concluded that all equilibrium toroidal solutions studied in that paper are dynamically stable to black hole collapse. This is not in disagreement with our conjecture and findings, since the solutions studied in \cite{Abrahams:1994kl} are not as relativistic as those presented here.

It is interesting to note that the $L_0 = 0.9$ sequence, which has a larger angular momentum, cannot be extended as far as the $L_0 = 0.8$ sequence. Beyond the terminal value of $E_0 = 0.628$, the iteration fails to converge. This could indicate another boundary of the solution space, similar to the mass-shedding limit observed in the uniformly rotating fluid case \cite{Meinel:2012tn}, and it would be interesting to pursue this question of why the iteration fails to converge further.

A related interesting open question is whether the sequences of relativistic and rotating toroidal solutions to the Einstein--Vlasov system, which we present here, exhibit a quasi-stationary transition to the extreme Kerr black hole. Such a transition has been observed in the case of uniformly rotating fluids \cite{Meinel:2012tn}. Evidence in favor of such a transition is provided by the approach of $\mathcal J/\mathcal M^2$ to $1$ as $E_0$ is decreased, although additional studies which push this sequence to lower $E_0$-values must be performed. Another line of support is the shape of the ergoregion, which in the limit is expected \cite{Meinel:2012tn} to form two lobes in the meridional plane which meet at the axis of rotation. Studies of this limiting behavior are ongoing.

We briefly comment on the numerical aspects of constructing these solution sequences. The solution mesh has a large radius  (compared to the support of the matter)  $r_b = 50$, which is highly refined near the coordinate origin. At each step of the sequence, the previous iterate is used as an initial guess in the solver. To save computational costs, the solutions with $E_0 > 0.75$ (to the right of the vertical dashed line in each panel of \Figref{fig:EVR_rel_torus_solchars}) are not fully converged; the purpose of these solutions is to obtain a suitable initial guess for the more relativistic solutions of greater interest. Solutions with $E_0 \leq 0.75$ have converged with a tolerance of $10^{-4}$. Additionally, the amount of damping in the fixed-point iteration is increased in three stages by decreasing from $\theta = 0.75$ to $\theta = 0.25$ during the sequence.

\begin{figure}[htb!]
\centering
  \centering
  \includegraphics[width=.7\linewidth]{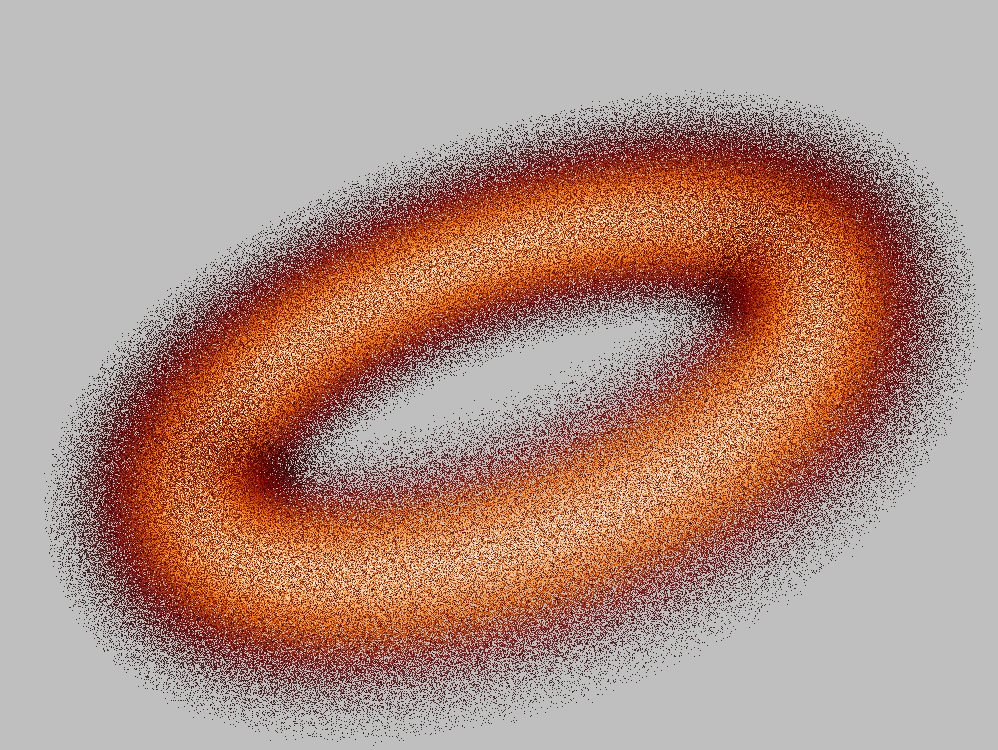}
  \captionsetup{width=.7\linewidth}
  \caption{A pointcloud visualization of the relativistic torus in the case $E_0 =0.58$. More dense regions are colored in white, while the less dense regions are displayed in darker shades.}
\label{fig:EVRRelTorusPointcloud}
\end{figure}

Finally, we remark that in earlier work \cite{Shapiro:1993hi} Shapiro and Teukolsky studied relativistic toroidal solutions using a delta-function ansatz. The most relativistic solution they were able to construct at the time had a parameter $R_0 = 4.5$, corresponding to $E_0 = 0.75$, a total angular momentum $\mathcal J = 1.34$, and did not contain an ergoregion. This is consistent with our results presented above, which indicate that ergoregions form at lower $E_0$-values. Interestingly, the authors note that beyond $R_0 = 4.5$ their iteration failed to converge. This is the same obstacle we encounter with the $L_0 = 0.9$ sequence discussed above, and it would be interesting to study if there is a common physical reason.

\subsection{Disk-Like Solutions}
\label{sec:Disks}
In this section we investigate solutions to the Vlasov-Poisson and Einstein-Vlasov systems with flattened spheroidal spatial density profiles, which can provide models for disk-like galaxies. 
While several authors have constructed disk-models in which the matter is confined to the plane \cite{Schenk:1999jv,Andreasson:2014gb}, our aim here is to find fully three-dimensional solutions whose spatial density distributions are as close to planar as possible.

\begin{figure}[htb!]
\centering
  \includegraphics[width=.5\linewidth]{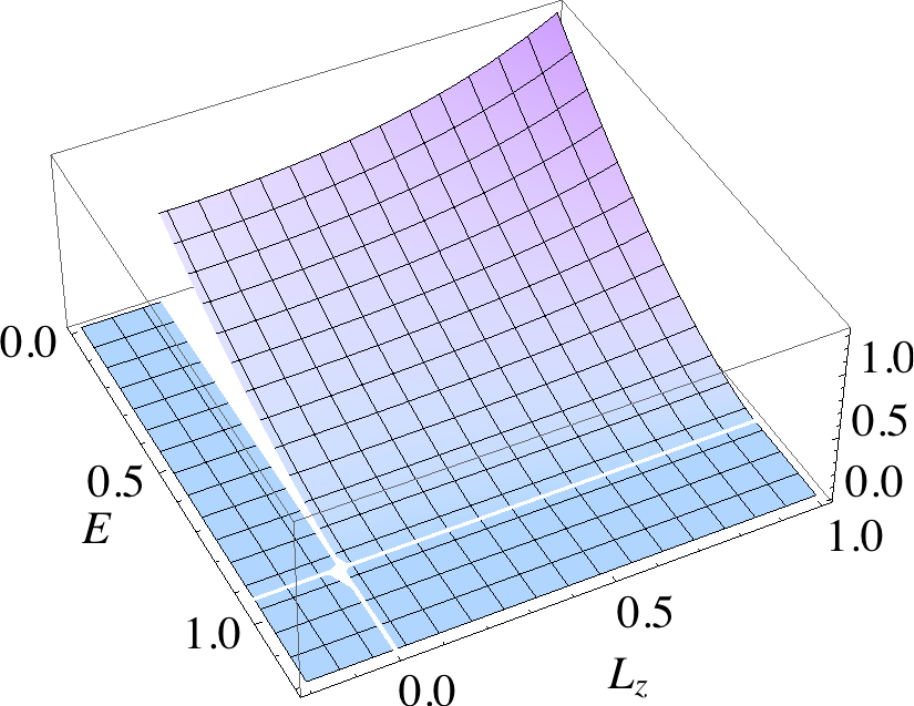}
  \captionsetup{width=.5\linewidth}
  \captionof{figure}{Ansatz function $\phi(E) \psi(L_z)$ given by \Eqref{eq:PolytropicEAnsatz} and \Eqref{eq:GaussianLAnsatz}, and with parameters chosen as in EVR solution of \Tableref{table:AlvarDiskCompare}. The ansatz generates solutions with net angular momentum since all particles are restricted to rotate with the same orientation.}
  \label{fig:evr_disk_ansatz}
\end{figure}

In our numerical experiments we find that the most flattened disks are generated by an ansatz having a Gaussian distribution in the angular momentum
\begin{equation}
\label{eq:GaussianLAnsatz}
\psi(L_z) = \frac 1L_0 \exp( L_z^2/L_0^2).
\end{equation}
In the limit $L_0 \to \infty$ the ansatz becomes independent of $L_z$, thus generating a spherically symmetric spatial density. As $L_0$ is decreased, particles with higher angular momentum are more heavily weighted compared to those with low angular momentum as shown in \Figref{fig:evr_disk_ansatz}. As before the distribution is taken to have a product structure with a polytropic distribution for $E$ as in \Eqref{eq:PolytropicEAnsatz}.
An ansatz of this type has been considered in \cite{Shapiro:1993gb} for creating spindle-type densities (by taking a negative sign in the exponential), and in \cite{Shapiro:1993hi} where rotating oblate solutions were presented. The same authors also investigated an ansatz of this type in the Newtonian case in \cite{Shapiro:1992ei}.

Despite being the most flattened solutions which we are able to produce, we have not been able to find solutions with spatial density configurations that approach an infinitely thin disk. 
For the present paper, we first illustrate, in the Einstein--Vlasov model, the extent to which we are able to obtain flattened solutions. Since the ansatz \Eqref{eq:GaussianLAnsatz} is even, these solutions have zero net angular momentum. We then compare disk-like solutions for the Vlasov--Poisson model (VP), and for the Einstein--Vlasov model both in the case above with zero net angular momentum (EV), and in the case where the ansatz  \Eqref{eq:GaussianLAnsatz} includes a momentum cutoff so that all particles are rotating in the same direction (EVR).

\begin{figure}[htb!]
\centering
  \includegraphics[width=.7\linewidth]{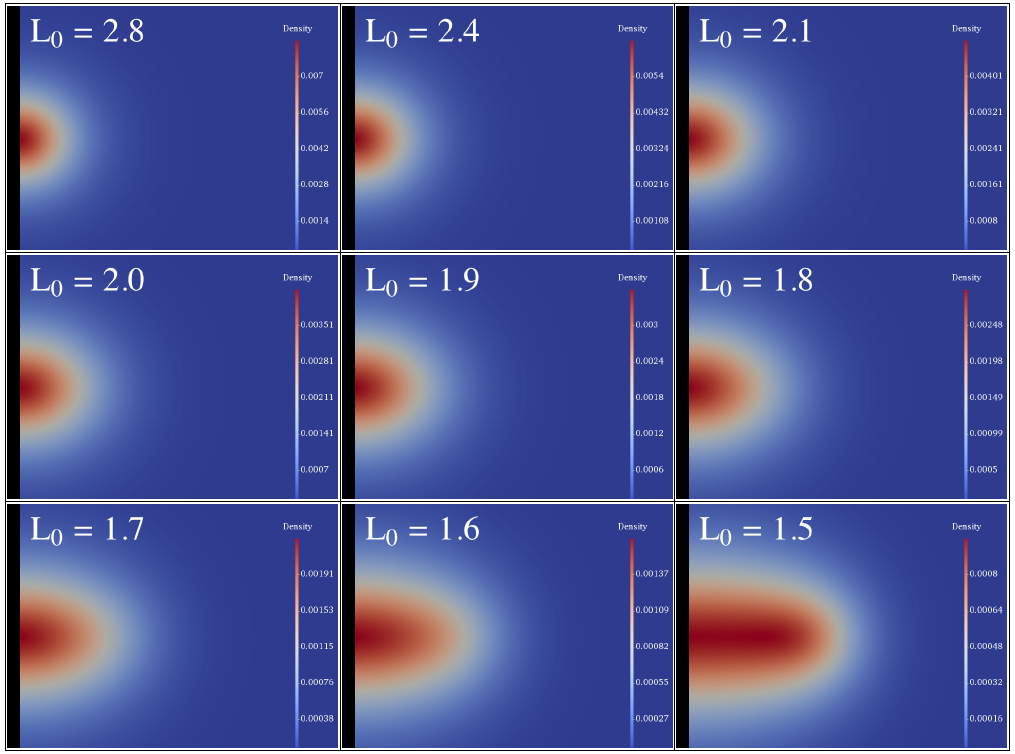}
\captionsetup{width=.7\linewidth}
  \caption{
  $L_0$-parameterized sequence of disk-like solutions in the Einstein--Vlasov model with zero net angular momentum $\mathcal J = 0$.
  }
  \label{fig:ev_disk_family}
\end{figure}

\Figref{fig:ev_disk_family} shows the density for a family of oblate spheroids with a Gaussian distribution in the angular momentum. The parameters are chosen $E_0 = 0.942, k = 1.5, l = 0$, and $L_0$ is decreased from $L_0 = 10$ to $L_0 = 1.5$. A deviation from spherical symmetry is only observed in the last portion of this sequence beginning around $L_0 \sim 2.8$. Within this parameter range the spatial density distribution stretches to its most flattened form, while for parameters $L_0 < 1.5$ the configuration appears not to remain gravitationally bound.

The minimum $L_0$ parameter for which the solution remains bounded depends, naturally, on the other parameters of the model. In particular, if the parameter $k$ is increased, leading to a more centrally condensed spatial density distribution, then the $L_0$ parameter can often be decreased further,  leading to a more flattened and disk-like distribution. Through such investigations we identify flattened configurations for this ansatz in each of the models. A table comparing the solution characteristics is presented in \Tableref{table:AlvarDiskCompare}. The $E_0$ parameters for these solutions were chosen such that the radius of support for both the relativistic solutions and the Newtonian solution were approximately equal.
\begin{table}[htp!]
\begin{center}
\begin{threeparttable}
    \footnotesize
    \begin{tabular}{lc|cc}
      \toprule
      \textbf{Model} & \textbf{Parameters} &  \textbf{Solution Characteristics} \\
      \midrule
      VP   	& $E_0 = -0.06$, $k=2.4$, $L_0 = 1.1$ 	& $w_{p} =2.14 $, $R_0 = 17.87$, $K^{-1} = 1.65$
                  \\[1em]
      EV   	& $E_0 = 0.942$, $k=2.0$, $L_0 = 1.40$  & $w_p = 1.55 $, $R_0 = 17.99$, $K^{-1} = 5.17$ , \\
      		&								&	$E_b = 0.032$,  $Z_c = 0.269$
                  \\[1em]
      EVR   & $E_0 = 0.942$, $k=1.6$, $L_0 = 1.27$	& $w_p = 0.87$,  $R_0 = 18.093$,  $K^{-1} = 4.90$, \\
      											& & $E_b = 0.029$, $Z_c = 0.216$, $\mathcal J = 1.1761$
                  \\[1em]
      \bottomrule
    \end{tabular}
    \normalsize
  \caption{
A comparison of disk-like solutions in the Vlasov--Poisson and Einstein--Vlasov models with and without net rotation. The peak density, $w_p$, is in units of $10^{-3}$. Note, that in the Vlasov--Poisson case (VP) we use the same symbol $w_p$ for the peak density, although in this case the density is obtained from the expression \Eqref{eq:VPDensityIntegral} for $w_0$.
}
\label{table:AlvarDiskCompare}
\end{threeparttable}
\end{center}
\end{table}

\begin{figure}[htb!]
  \begin{center}
  \begin{subfigure}[b]{0.3\linewidth}
    \centering
        \includegraphics[width=0.95\linewidth]{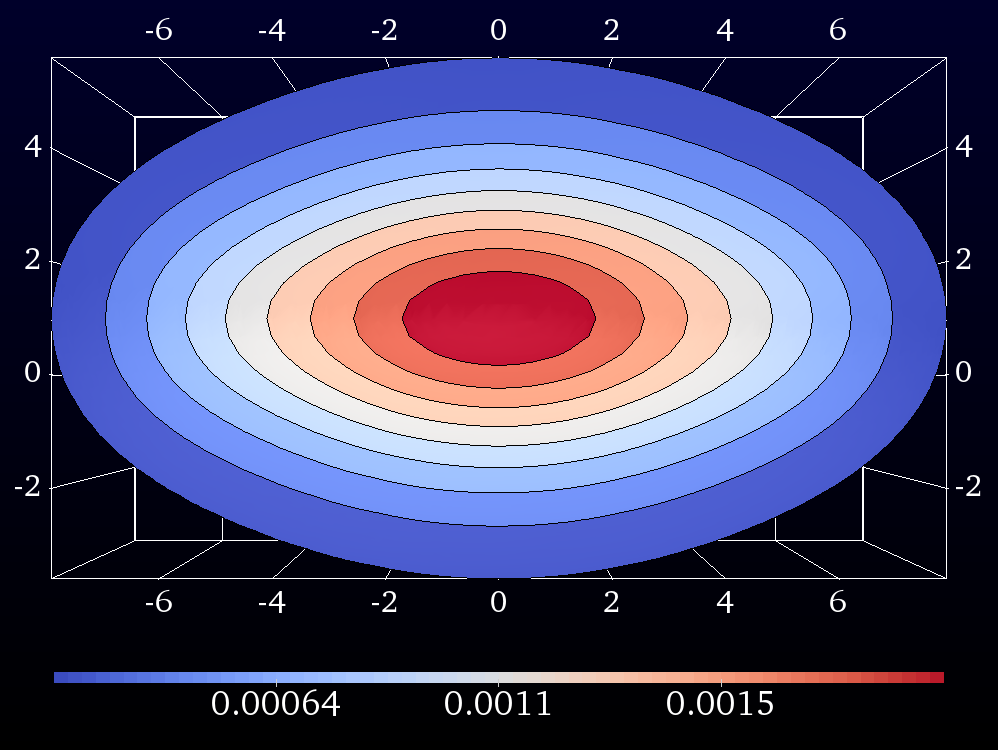}
    \caption{}
  \end{subfigure}
  \begin{subfigure}[b]{0.3\linewidth}
    \centering
    \includegraphics[width=0.95\linewidth]{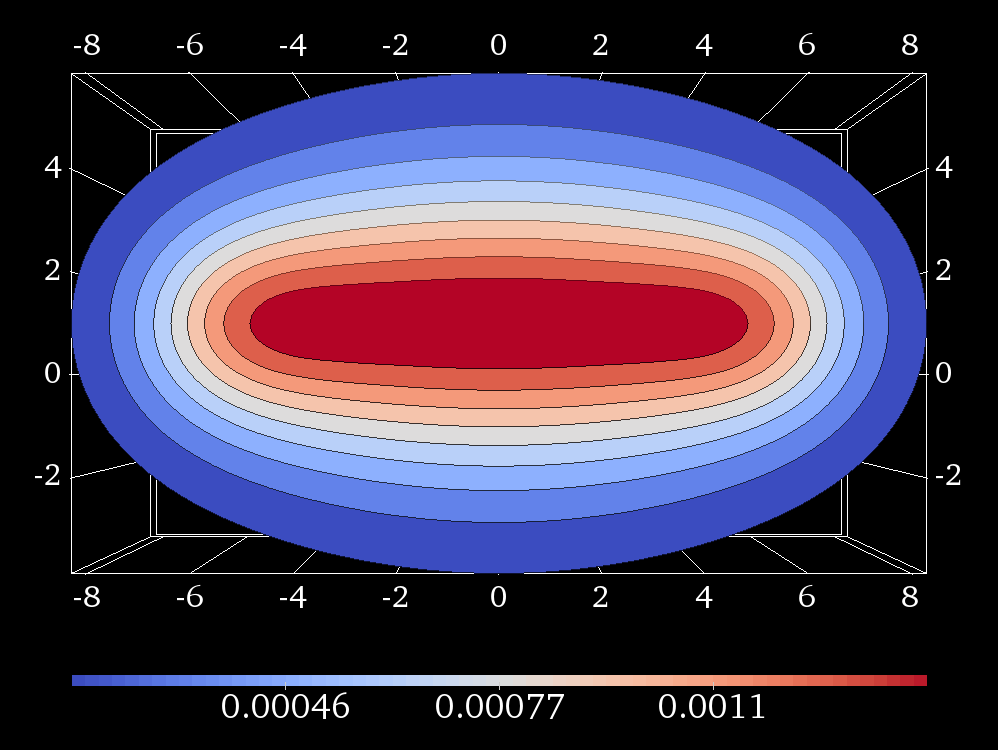}
    \caption{}
  \end{subfigure}
  \begin{subfigure}[b]{0.3\linewidth}
    \centering
        \includegraphics[width=0.95\linewidth]{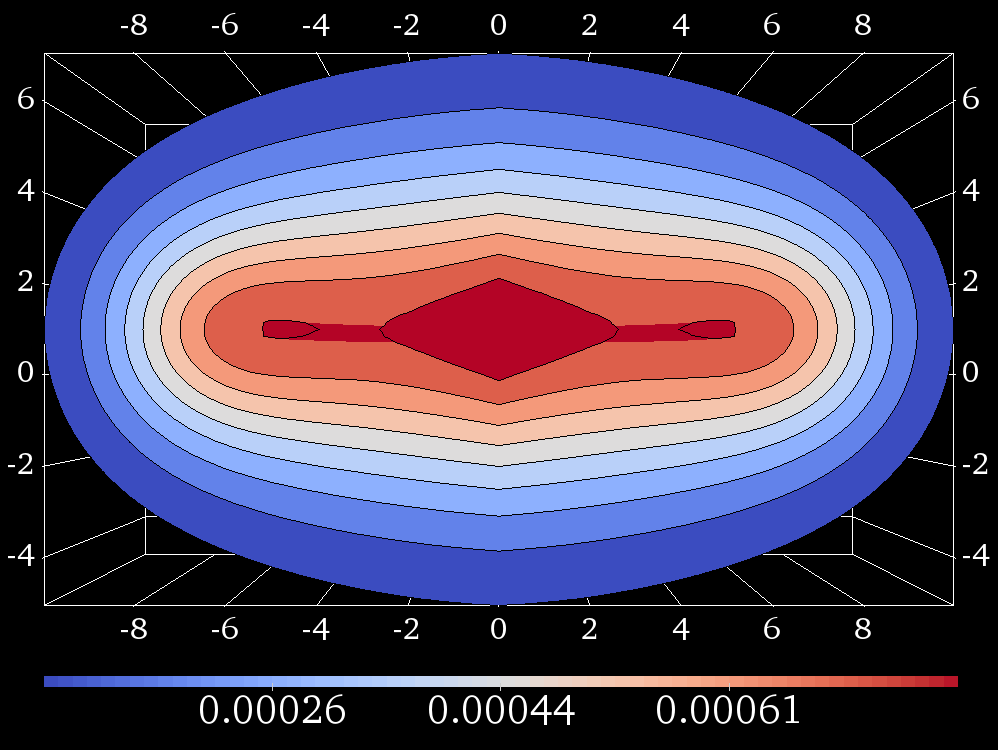}
    \caption{}
  \end{subfigure}
  \captionsetup{width=0.9\linewidth}
    \caption{Panels (a)--(c) show density contours for values $10\%$ through $90\%$ of peak density for the models VP, EV, and EVR respectively with parameters listed in \Tableref{table:AlvarDiskCompare}.
    }
  \label{fig:DiskContours}
  \end{center}
\end{figure}

\begin{figure}[htb!]
\centering
  \begin{subfigure}[b]{0.35\linewidth}
    \centering
    \includegraphics[width=0.95\linewidth]{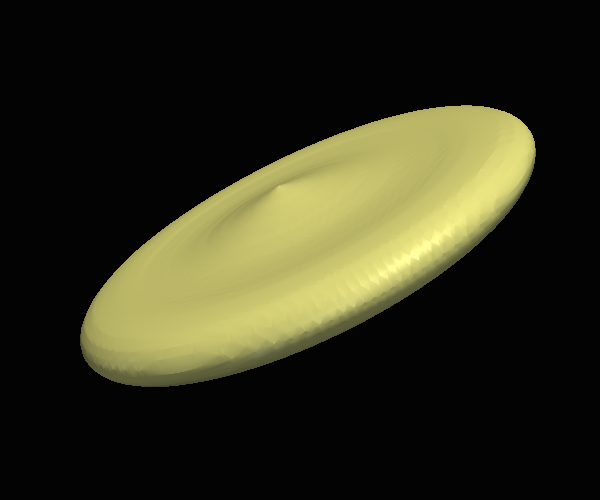}
        \caption{}
  \end{subfigure}
  \begin{subfigure}[b]{0.35\linewidth}
    \centering
    \includegraphics[width=0.95\linewidth]{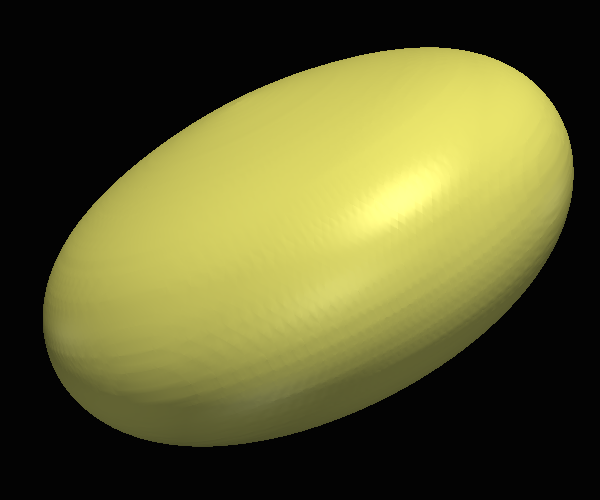}
        \caption{}
  \end{subfigure}
    \captionsetup{width=0.7\linewidth}
  \caption{Two density contours for the disk-like solution in the EVR model (see \Tableref{table:AlvarDiskCompare}). Panels (a), (b) show the contours at $85\%$ and $20\%$ of peak density respectively.}
  \label{fig:TwoDiskDensityContours}
\end{figure}

Despite the similarity in the solution characteristics presented in \Tableref{table:AlvarDiskCompare}, there are differences in the character of the solutions in each case. As shown in the spatial density contour plots \Figref{fig:DiskContours}, at low densities the solutions are very similar, while at higher densities the solutions to the Einstein--Vlasov equations have more flattened contours (see also \Figref{fig:TwoDiskDensityContours}). We note that none of the solutions are particularly relativistic in terms of the parameter $2 \mathcal M /R_0 \sim 0.1$.

At higher density contours the rotating relativistic solution also displays a central bulge and a toroidal region. The peak density is at the origin, while the contours are toroidal only for densities close to ninety percent of peak. We note that solutions obtained via a similar ansatz were studied in \cite{Shapiro:1993hi}. In that study the authors present a family of solutions with a fixed polytropic exponent (here called $k$), and varied $L_0$. For small $L_0$ values, they find that the peak density occurs in a ring, rather than at the center. In fact, the solutions presented in \cite{Shapiro:1993hi} contain no central bulge. This difference in the character of the solutions is due to the polytropic ansatz, which in \cite{Shapiro:1993hi} is chosen such that particles of all energies are weighted equally. In contrast, as shown in \Figref{fig:evr_disk_ansatz}, our ansatz suppresses higher energy particles.
We also remark that one can obtain toroidal like structures even in the Vlasov--Poisson model. Indeed, solutions obtained in \cite{Shapiro:1992ei} exhibit such structure, where the choice of parameters in that paper corresponds to taking the polytropic exponent $k<0$. However, since these solutions are less flattened, we have not presented them here.

Solutions to the Vlasov-Poisson system have been shown to be useful models in astrophysics; see for example \cite{binney2011galactic} and references therein. It would be very interesting to extend our study with the aim of accurately modeling galaxies. Such studies could include other observable characteristics of galaxies, such as velocity dispersion profiles and rotation curves, and also dark matter components using multiple ansatz functions as in \Sectionref{sec:CompositeObjects}.

\subsection{Spindle Solutions}
\label{sec:SpindleSolutions}
As a further test of our code and demonstration of different ansatz functions, we present two solutions with spindle-like spatial density distributions.

\begin{figure}[h]
\centering
  \begin{subfigure}[b]{0.35\linewidth}
    \centering
    \includegraphics[width=0.95\linewidth]{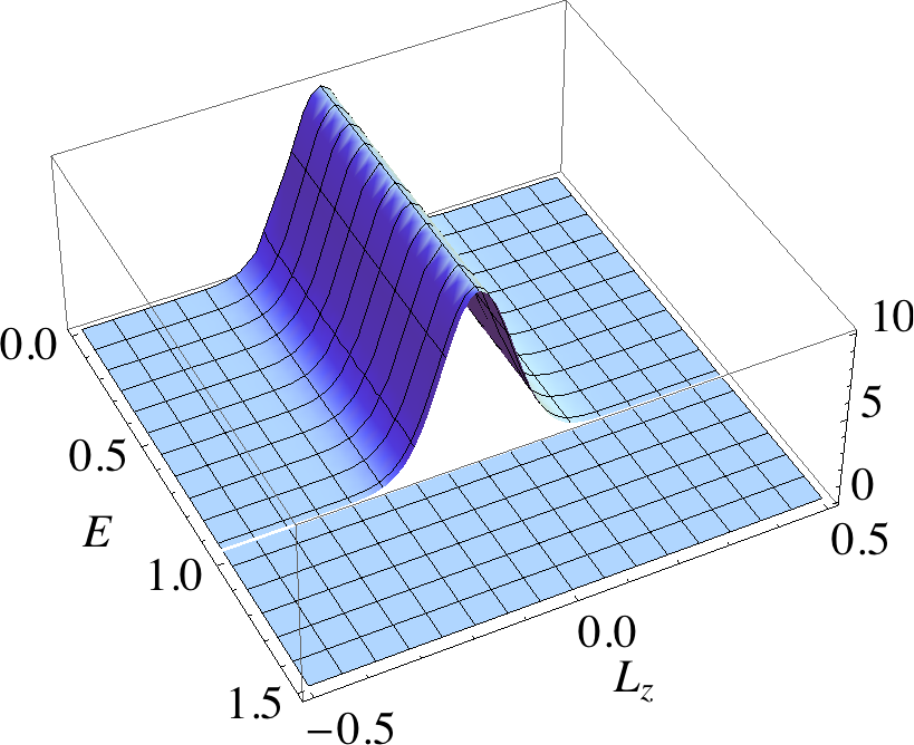}
        \caption{}
  \end{subfigure}
  \quad   \quad
  \begin{subfigure}[b]{0.35\linewidth}
    \centering
    \includegraphics[width=0.95\linewidth]{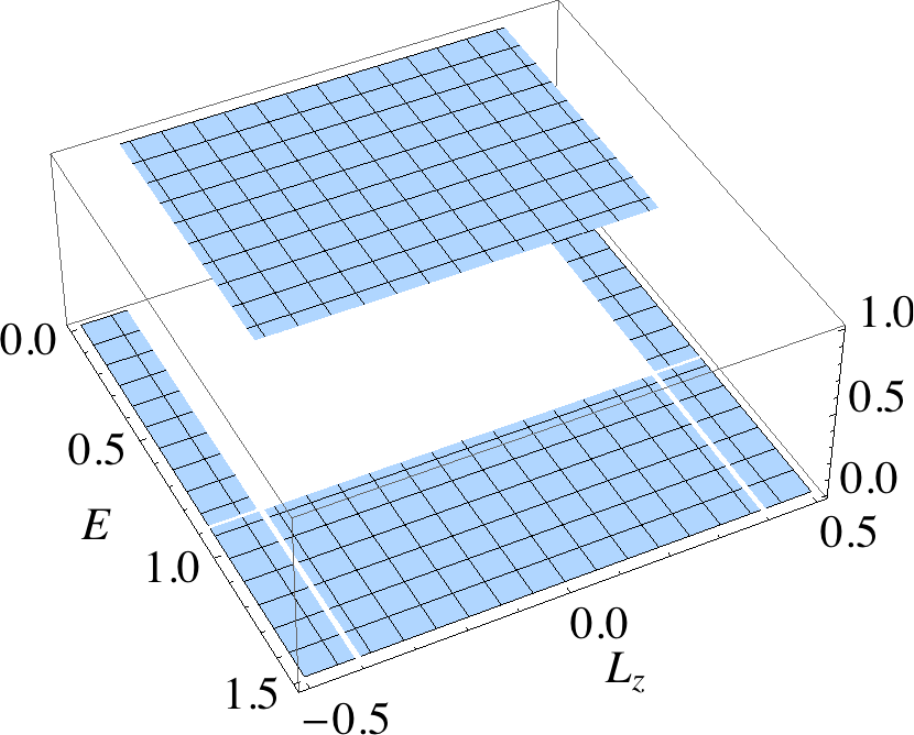}
        \caption{}
  \end{subfigure}
\captionsetup{width=0.7\linewidth}
\caption{Ansatz functions for spindle solutions. The Gaussian ansatz is in Panel (a), and the Polytropic--Spindle ansatz is shown in Panel (b).}
  \label{fig:SpindleAnsatzFunctions}
\end{figure}

The first of these solutions is based on a Gaussian distribution in angular momentum \Eqref{eq:GaussianLAnsatz}, but with a negative sign in the exponential; see \Figref{fig:SpindleAnsatzFunctions}, Panel (a). A sequence of stationary solutions with this ansatz has been studied by Shapiro and Teukolsky in connection with the formation of naked singularities \cite{Shapiro:1992ei,Shapiro:1993gb,Shapiro:1991kg}. Spindle configurations can also be constructed with a distribution in momentum of the form
\begin{equation}
\label{eq:AndreassonLAnsatz}
\psi(L_z) =
\begin{cases}
	(1 - Q |L_z| )^l, & |L_z| < 1/Q \\
       0,  		     & |L_z| \geq 1/Q ,\\
   \end{cases}
\end{equation}
which is illustrated \Figref{fig:SpindleAnsatzFunctions}, Panel (b). This ansatz enforces an upper bound on $L_z$, controlled by $Q>0$, which in turn forces particles to be close the axis. We refer to this choice as the \emph{polytropic--spindle} ansatz. For both of these ansatzes we use a polytropic distribution in the particle energy \Eqref{eq:PolytropicEAnsatz}.
We compute both of these solutions in the Einstein--Vlasov case with equal numbers of particles rotating in both directions. Similar solutions can be obtained with particles rotating in only one direction, as well as in the Vlasov--Poisson case.

\begin{figure}[htp!]
\centering
    \includegraphics[width=0.35\linewidth]{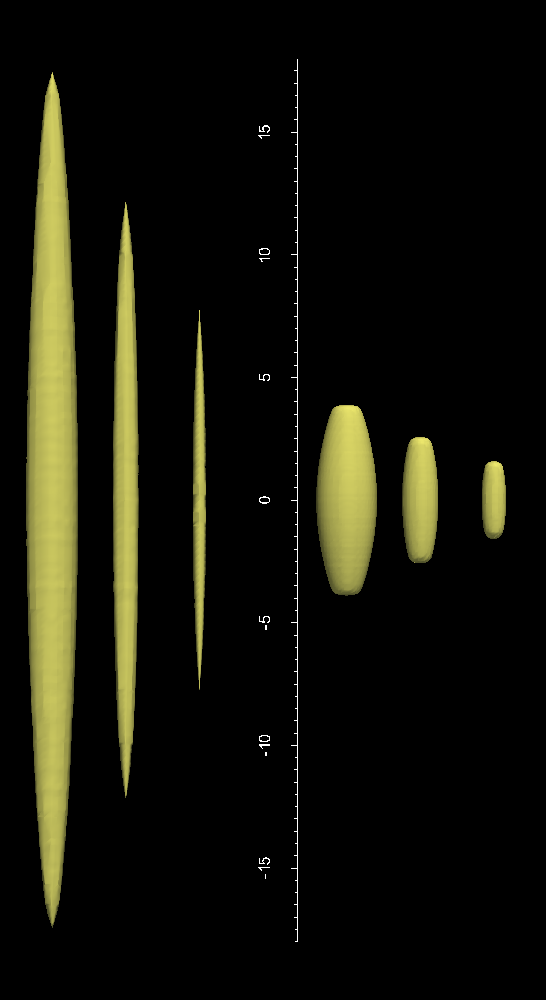}
\captionsetup{width=0.35\linewidth}
\caption{Contours at $25 \%$, $50 \%$, and $75 \%$ of peak density from two spindle solutions are displayed. Left of the axis are the contours of a spindle solution from the Gaussian ansatz, and to the right of the axis are contours of the Polytropic-Spindle ansatz solution. The axis in this figure gives a measure of the coordinate extent of the contours, but does not represent the axis of symmetry for the solutions.}
  \label{fig:SpindleContours}
\end{figure}

The parameters and characteristics of the solutions are shown in \Tableref{table:SpindleSolutions}. In both solutions the peak density occurs at the coordinate origin, while as shown in \Figref{fig:SpindleContours}, the Gaussian spindle solution has particles which cluster more strongly on the axis and have a more pronounced spindle shape.

\begin{table}
\begin{center}
\begin{threeparttable}
    \footnotesize
       \begin{tabular}{lc|cccccc}
      \toprule
      \textbf{$L_z$-Ansatz} & \textbf{Solution Parameters}  &  \multicolumn{6}{c}{\textbf{Solution Characteristics}}   \\
       		    		& 			   			& $w_p$ & $R_0$ & $2\mathcal{M}/R_0$ & $E_b$ & $Z_c$  & $K^{-1}$  \\
            \midrule
Gaussian &  $E_0 = 0.966$, $k=0.0$, $L_0 = 0.1$ 	& $0.002$ & $32.93$ & $0.06$ & $ 0.016$ & $0.121$ & $ 2235.5$ \\
 		[1em]
Polytropic-Spindle &  $E_0 = 0.9$, $Q = 2.5$, $k=l = 0$ & $0.02$ & $11.18$ &  $ 0.18$ &  $ 0.035$ &  $ 0.477$ & $ 449.27$ \\
                  [1em]
      \bottomrule
    \end{tabular}
  \caption{
	Parameters and characteristics for two spindle solutions. The peak density is denoted $w_p$, $K$ is the ansatz coefficient, $E_b$ is the fractional binding energy, $Z_c$ is the central redshift, and $R_0$ is the radius of support measured in the R-coordinate cf.~\Eqref{eq:RCoord} of \Sectionref{sec:SolutionCharacteristics}.
  }
  \label{table:SpindleSolutions}
  \end{threeparttable}
  \end{center}
\end{table}

The parameter $E_0 = 0.966$ (corresponding to $R_0 = 30$, cf.~\Eqref{eq:E0andR0} above) is chosen to agree with the parameter for the most extreme polytropic spindle solution in \cite{Shapiro:1993gb} (see Table 2). We note that we are able to reproduce the solutions with $R_0 = 30$ in that table. The solution presented here however makes the ``democratic" choice $k =0.0$, giving equal weight to particles of all energies, while the solutions in Table 2 of \cite{Shapiro:1993gb} use $k =1.0$.

\subsection{Composite Spindle-Torus Objects}
\label{sec:CompositeObjects}
One of the strengths of our code is the ability to quickly implement new ansatz functions and to treat composite models formed by summing together multiple ansatz functions. Examples of composite astrophysical objects are numerous, and include disk galaxies with a central bulge, galaxies with dark matter halos, and ring-type galaxies.

There are examples of composite models in the Newtonian case existing in the literature. Fricke \cite{Fricke:1952vb} expands the distribution in terms of the form $(E_0-E)^kL_z^{2n}$ for integers $n$. Toomre \cite{Toomre:1982aa} considers distributions of the form $L_z^{2n} e^{-E/\sigma^2}$, which have vanishing density at the axis for non-zero $n$, and increasingly flattened peanut-shaped projections in the meridional plane for $n\ge 1$. Such models are combined to construct central bulge-disk and halo-disk configurations. Later, Evans \cite{Evans:vc} shows that an axisymmetric logarithmic potential of Binney \cite{Binney:1981aa} can be constructed from the sum of three of Toomre's components. This result is then used in constructing composite models with central stellar densities and dark halos. The existence of flat stellar disks confined to a plane with dark matter halos is proved in the work of Fi\v{r}t et al. \cite{Firt:2009hv}. Composite models allow for much more complexity in the density distributions, and greatly enlarges the space of solutions. We demonstrate the capability of our code to handle multiple distributions by presenting a two-component family of spindle-torus objects, which may provide models for ring-type galaxies. While the above works are done in the Vlasov--Poisson model, to the authors' knowledge the solutions obtained here are the first example of composite objects studied in the Einstein--Vlasov system. Similar solutions may also be computed in the Vlasov--Poisson model.

The composite ansatz is taken to have the form
\[\Phi(E, L_z) = C_s \Phi_{\mathrm{spindle}}(E,L_z) + C_t \Phi_{\mathrm{torus}}(E,L_z),\]
where $\Phi_{\mathrm{spindle}}(E,L_z)$ uses the polytropic-spindle ansatz introduced above \Eqref{eq:AndreassonLAnsatz}, and  $\Phi_{\mathrm{torus}}(E,L_z)$ uses a polytropic type ansatz with nonzero $L_0$ (cf.~\Sectionref{sec:PolytropicAnsatzSolutions}).
It is interesting to note that we were not successful in combining any ansatz for the central object with a torus. Our initial attempts of combining a polytropic central bulge with torus resulted in either a central bulge or a torus, and both configurations only occurred simultaneously with significant overlap.

\begin{table}[htp!]
  \begin{center}
\begin{threeparttable}
    \footnotesize
    \begin{tabular}{l|cccccccc}
      \toprule
     \textbf{Solution Parameter} &  \multicolumn{8}{c}{\textbf{Solution Characteristics}}   \\
      $L_0$	 &   $w_c$ ($\times 10^{-4}$) & $w_c/w_v$ & $w_c/w_t$ & $\rho_v$ & $\rho_t$ & $Z_c$ & $ E_b$ & $R_0$ \\
            \midrule
      $1.3$ &   $1.6$ & $1.7$ & $0.2$ & $4.0$ & $8.8$ & $0.13$ & $0.022$ & $18.2$ \\
      $1.4$ &   $1.6$ & $1.9$ & $0.3$ & $4.3$ & $8.8$ & $0.13$ & $0.022$ & $18.0$ \\
      $1.5$ &   $3.0$ & $2.3$ & $0.5$ & $4.3$ & $8.6$ & $0.14$ & $0.022$ & $17.9$ \\
      $1.6$ &   $6.5$ & $3.3$ & $1.1$ & $4.3$ & $8.0$ & $0.17$ & $0.023$ & $17.7$ \\
      $1.7$ &   $6.5$ & $5.2$ & $3.5$ & $4.0$ & $6.7$ & $0.22$ & $0.025$ & $17.4$ \\
      $1.8$ &   $41.0$ & $8.8$ & $8.6$ & $4.6$ & $5.2$ & $0.28$ & $0.029$ & $17.1$   \\
      \bottomrule
    \end{tabular}
\caption{
Family of spindle-torus objects where the $L_0$-parameter for the torus distribution is varied. The constants are taken to be $C_s = 0.5$, $C_t = 1.0$. The central density $w_c$ is in units of $10^{-4}$. The total mass energy of each of the solutions is taken to be $\mathcal M = 1$. The central, valley, and peak of torus densities are labeled $w_c, w_v, w_t$ respectively. We let $\rho_v$ and $\rho_t$ denote the coordinate radius of the valley and peak of ring.
}
\label{table:shoFamilyTable}
\end{threeparttable}
  \end{center}
\end{table}

\begin{figure}[htp!]
\centering
  \begin{subfigure}[b]{0.3\linewidth}
  \centering
  \includegraphics[width=.9\linewidth]{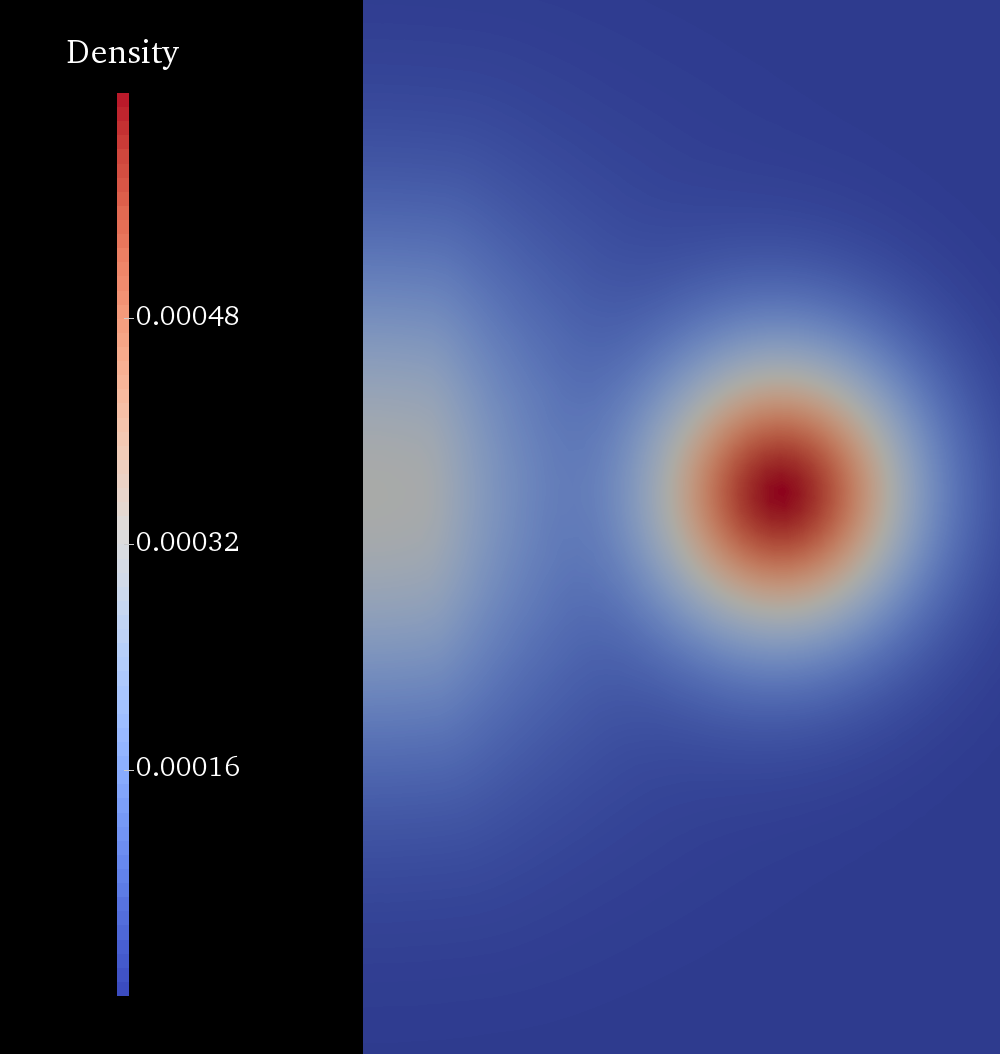}
  \caption{$L_0 = 1.5$}
  \label{fig:shofamily150}
  \end{subfigure}
  \begin{subfigure}[b]{0.3\linewidth}
  \centering
  \includegraphics[width=.9\linewidth]{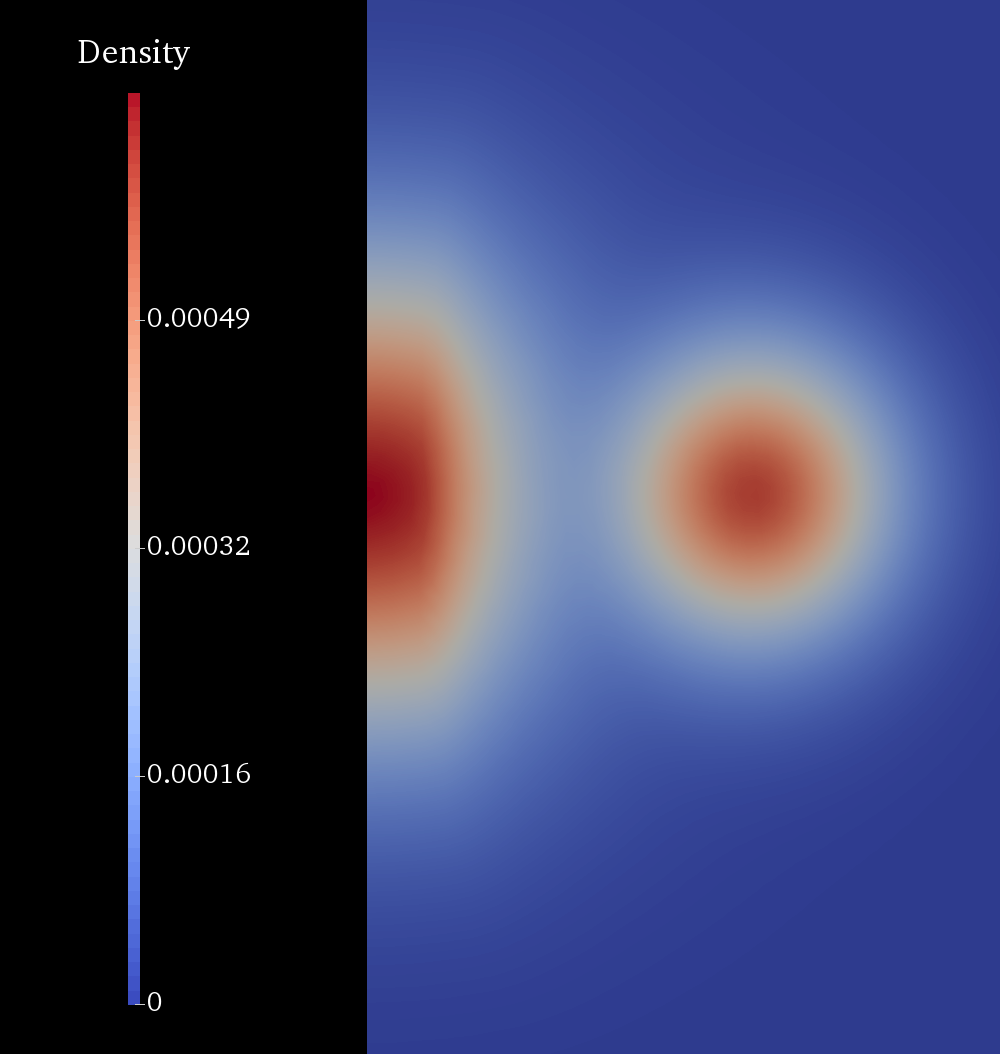}
  \caption{$L_0 = 1.6$}
  \label{fig:shofamily160}
  \end{subfigure}
  \begin{subfigure}[b]{0.3\linewidth}
  \centering
  \includegraphics[width=.9\linewidth]{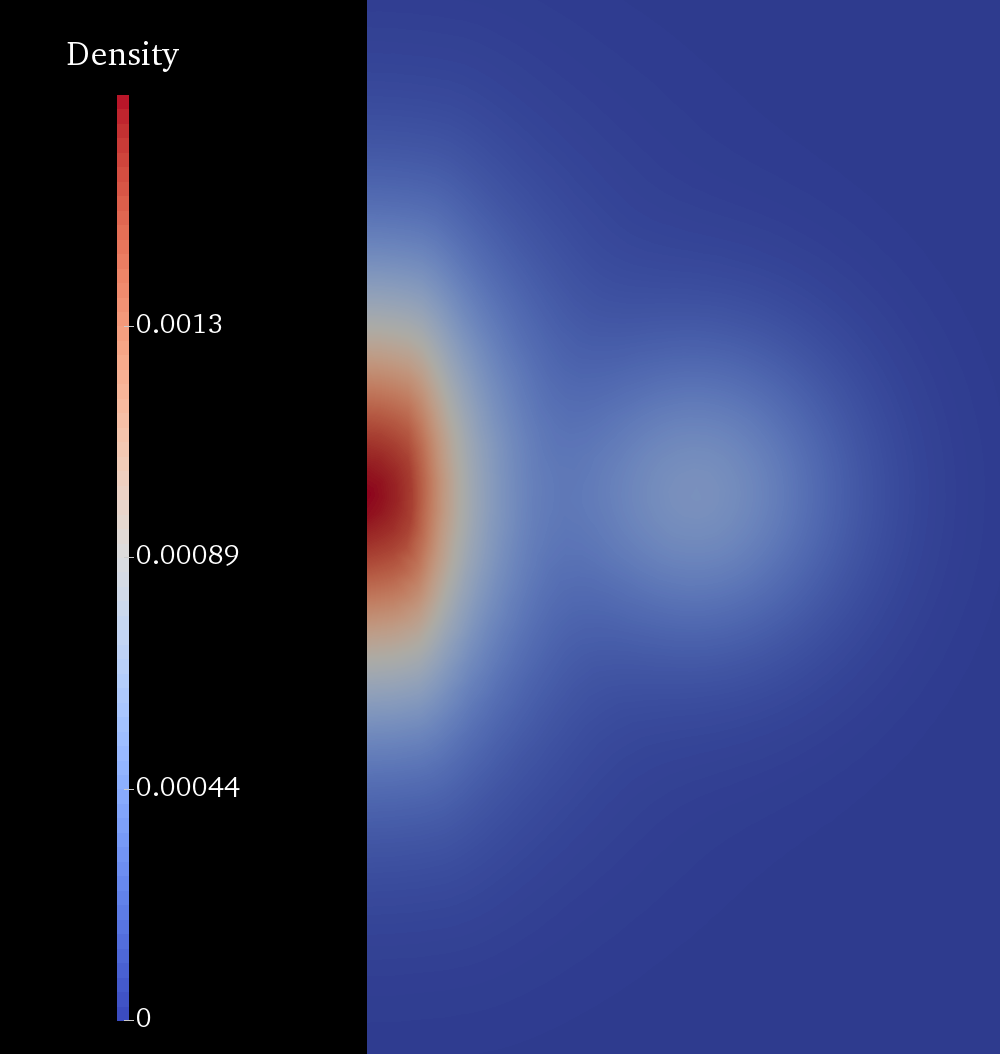}
  \caption{$L_0 = 1.7$}
  \label{fig:shofamily170}
  \end{subfigure}
  \captionsetup{width=0.9\linewidth}
  \caption{Energy densities for three members of the family of spindle-torus objects in \Tableref{table:shoFamilyTable}. In (a) $L_0 = 1.5$, in (b) $L_0 = 1.6$, and in (c) $L_0 = 1.7$. }
\label{fig:Spindle_Torus_Family}
\end{figure}

In \Tableref{table:shoFamilyTable} we exhibit members of a family of solutions parametrized by the $L_0$ parameter for the torus component. These solutions are computed using the Einstein--Vlasov solver, although none of the solutions are relativistic in the sense of a high $2\mathcal M/R_0$ value. The solutions shown here have zero net angular momentum. For this simulation $E_0 = 0.940$. The parameters for the spindle ansatz are fixed to be $C_s = 0.5$, $Q = 2, k = 1, l = 0$, while for the torus component we take $k=1$, $l=1$ and vary $L_0$ between $1.8$ and $1.3$. Outside of this range, the density resides nearly entirely in one of the components. Density profiles for three selections from the family are presented in \Figsref{fig:shofamily150}---\ref{fig:shofamily170}. Although the solutions in \Figref{fig:Spindle_Torus_Family} exhibit a near-vacuum region between the components, we have not been able to construct solutions which have a complete vacuum in this region. In \Tableref{table:shoFamilyTable} we list the ratios of the central density to the density in the valley between components, and the central density to the peak torus density. We note that the $L_0 = 1.6$ solution has nearly equal central and torus-peak densities, and for this case the valley density is approximately 3 times less dense. It is very likely that by further exploring the large parameter space one can find solutions with a more pronounced vacuum region separating the components.

\begin{figure}[htb!]
\centering
  \includegraphics[width=.7\linewidth]{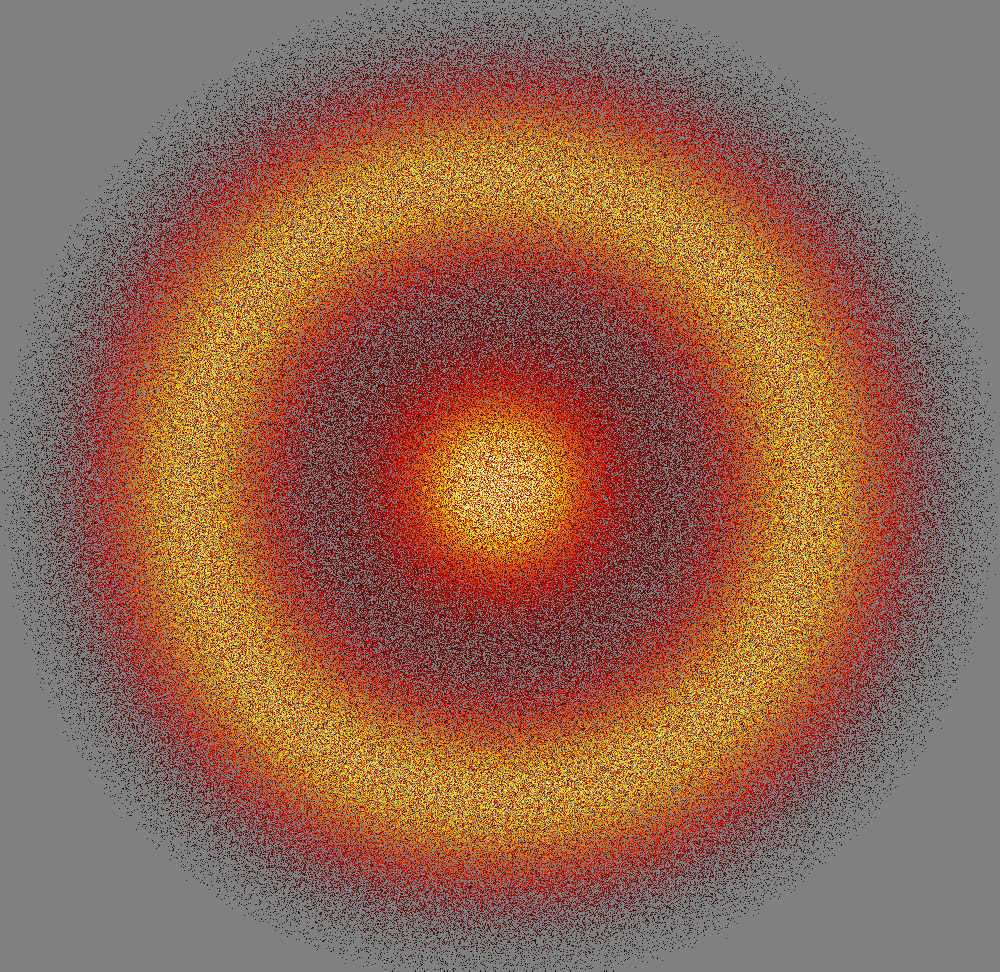}
  \captionsetup{width=.7\linewidth}
  \caption{Pointcloud representation of the $L_0 =1.6$ spindle-torus solution viewed from along the axis of symmetry.}
  \label{fig:spindlehoag_fancy}
\end{figure}

We remark that astrophysical objects of this form occur in nature, for instance Hoag's object \cite{HOAG:1950vn}, and other ring-type galaxies. A three-dimensional pointcloud representation of the $L_0 = 1.6$ solution discussed above is shown in \Figref{fig:spindlehoag_fancy}.

\section{Acknowledgments}
The authors thank Lars Andersson, Marcus Ansorg, Reinhard Meinel, and Gerhard Rein for comments and helpful discussions during the preparation of this manuscript.

\small{
\bibliography{bibliography}

\providecommand{\newblock}{}
\begin{thebibliography}{10}
\expandafter\ifx\csname url\endcsname\relax
  \def\url#1{{\tt #1}}\fi
\expandafter\ifx\csname urlprefix\endcsname\relax\def\urlprefix{URL }\fi
\providecommand{\eprint}[2][]{\url{#2}}

\bibitem{binney2011galactic}
Binney J and Tremaine S 2011 {\em {Galactic dynamics}\/} (Princeton university
  press)

\bibitem{Rein:ud}
Rein G 2007 {\em Handbook of Differential Equations\/}
  \urlprefix\url{http://www.neu.uni-bayreuth.de/de/Uni_Bayreuth/Fakultaeten/1_Mathematik_Physik_und_Informatik/Mathematisches_Institut/mathe_VI-Rein/de/download/kinetic_elsevier.pdf}

\bibitem{Andreasson:2011dza}
Andr{\'e}asson H 2011 {\em Living Reviews in Relativity\/} {\bf 14}
  \urlprefix\url{http://www.livingreviews.org/lrr-2011-4}

\bibitem{Shapiro:1992ei}
Shapiro S~L and Teukolsky S~A 1992 {\em The Astrophysical Journal\/} {\bf 388}
  287--300 \urlprefix\url{http://adsabs.harvard.edu/doi/10.1086/171152}

\bibitem{Shapiro:1993gb}
Shapiro S~L and Teukolsky S~A 1993 {\em The Astrophysical Journal\/} {\bf 419}
  622--635 \urlprefix\url{http://adsabs.harvard.edu/doi/10.1086/173513}

\bibitem{Shapiro:1993hi}
Shapiro S~L and Teukolsky S~A 1993 {\em The Astrophysical Journal\/} {\bf 419}
  636--647 \urlprefix\url{http://adsabs.harvard.edu/doi/10.1086/173514}

\bibitem{Rein:2003cg}
Rein G and Guo Y 2003 {\em Monthly Notices of the Royal Astronomical Society\/}
  {\bf 344} 1296--1306
  \urlprefix\url{http://mnras.oxfordjournals.org/cgi/doi/10.1046/j.1365-8711.2003.06920.x}

\bibitem{Andreasson:2011hg}
Andr{\'e}asson H, Kunze M and Rein G 2011 {\em Communications in Mathematical
  Physics\/} {\bf 308} 23--47
  \urlprefix\url{http://link.springer.com/10.1007/s00220-011-1324-8}

\bibitem{Andreasson:ch}
Andr{\'e}asson H, Kunze M and Rein G 2014 {\em Communications in Mathematical
  Physics\/}
  \urlprefix\url{http://link.springer.com/article/10.1007/s00220-014-1904-5}

\bibitem{Bardeen:1973ux}
Bardeen J~M 1973 {Rapidly rotating stars, disks, and black holes} {\em Black
  Holes (Les Astres Occlus)\/} ed Dewitt C and Dewitt B~S pp 241--289
  \urlprefix\url{http://books.google.com/books?hl=en&lr=&id=16FpuO6h3A4C&oi=fnd&pg=PA241&dq=bardeen+rapidly&ots=C58OIy_Vk4&sig=xXmDeXtFbM_noBsS6B8HrS8D0OI}

\bibitem{Ansorg:2008jr}
Ansorg M and Pfister H 2008 {\em Classical And Quantum Gravity\/} {\bf 25}
  035009
  \urlprefix\url{http://stacks.iop.org/0264-9381/25/i=3/a=035009?key=crossref.a25cdb470d9a3bdca3f02a13dee3798e}

\bibitem{PhysRev.113.934}
Komar A 1959 {\em Phys. Rev. (2)\/} {\bf 113} 934--936
  \urlprefix\url{http://link.aps.org/doi/10.1103/PhysRev.113.934}

\bibitem{Beig:1978vr}
Beig R 1978 {\em Physics Letters A\/} {\bf 69} 153--155
  \urlprefix\url{http://www.sciencedirect.com/science/article/pii/0375960178901986}

\bibitem{ChoquetBruhat:2008te}
Choquet-Bruhat Y 2008 {\em {General Relativity and the Einstein Equations}\/}
  Oxford mathematical monographs (Oxford, New York: Oxford University Press)

\bibitem{Andreasson:2006dza}
Andr{\'e}asson H and Rein G 2006 {\em Classical And Quantum Gravity\/} {\bf 23}
  3659
  \urlprefix\url{http://iopscience.iop.org/article/10.1088/0264-9381/23/11/001}

\bibitem{Buchdahl:1959be}
Buchdahl H~A 1959 {\em Physical Review\/} {\bf 116} 1027--1034
  \urlprefix\url{http://link.aps.org/doi/10.1103/PhysRev.116.1027}

\bibitem{Andreasson:2008fu}
Andr{\'e}asson H 2008 {\em Journal of Differential Equations\/} {\bf 245}
  2243--2266
  \urlprefix\url{http://www.sciencedirect.com/science/article/pii/S0022039608002398}

\bibitem{Andreasson:2007kv}
Andr{\'e}asson H 2007 {\em Communications in Mathematical Physics\/} {\bf 274}
  409--425 \urlprefix\url{http://link.springer.com/10.1007/s00220-007-0285-4}

\bibitem{Brenner:2008hf}
Brenner S~C and Scott L~R 2008 {\em {The mathematical theory of finite element
  methods}\/} 3rd ed ({\em Texts in Applied Mathematics\/} vol~15) (Springer,
  New York) ISBN 978-0-387-75933-3
  \urlprefix\url{http://dx.doi.org/10.1007/978-0-387-75934-0}

\bibitem{Logg:2012jw}
Logg A, Mardal K~A and Wells G (eds) 2012 {\em {Automated Solution of
  Differential Equations by the Finite Element Method}\/} ({\em Lecture Notes
  in Computational Science and Engineering\/} vol~84) (Berlin, Heidelberg:
  Springer Berlin Heidelberg) ISBN 978-3-642-23099-8
  \urlprefix\url{http://link.springer.com/10.1007/978-3-642-23099-8}

\bibitem{Logg:2010kt}
Logg A and Wells G~N 2010 {\em ACM Transactions on Mathematical Software\/}
  {\bf 37} 1--28
  \urlprefix\url{http://portal.acm.org/citation.cfm?doid=1731022.1731030}

\bibitem{Andreasson:2014gb}
Andr{\'e}asson H and Rein G 2014 {\em Monthly Notices of the Royal Astronomical
  Society\/} {\bf 446} 3932--3942
  \urlprefix\url{http://mnras.oxfordjournals.org/cgi/doi/10.1093/mnras/stu2346}

\bibitem{Fischer:2005bw}
Fischer T, Horatschek S and Ansorg M 2005 {\em Monthly Notices of the Royal
  Astronomical Society\/} {\bf 364} 943--947
  \urlprefix\url{http://mnras.oxfordjournals.org/cgi/doi/10.1111/j.1365-2966.2005.09629.x}

\bibitem{Ansorg:2003dk}
Ansorg M, Kleinw{\"a}chter A and Meinel R 2003 {\em The Astrophysical
  Journal\/} {\bf 582} L87--L90
  \urlprefix\url{http://iopscience.iop.org/article/10.1086/367632}

\bibitem{Schbel:2003em}
Sch{\"o}bel K and Ansorg M 2003 {\em Astronomy and Astrophysics\/} {\bf 405}
  405--408
  \urlprefix\url{http://www.edpsciences.org/10.1051/0004-6361:20030634}

\bibitem{Abrahams:1994kl}
Abrahams A, Cook G, Shapiro S and Teukolsky S 1994 {\em Physical review D\/}
  {\bf 49} 5153--5164
  \urlprefix\url{http://link.aps.org/doi/10.1103/PhysRevD.49.5153}

\bibitem{Meinel:2012tn}
Meinel R, Ansorg M, Kleinw{\"a}chter A, Neugebauer G and Petroff D 2012 {\em
  {Relativistic Figures of Equilibrium}\/} (Cambridge University Press) ISBN
  9781107407350
  \urlprefix\url{http://books.google.se/books?id=-MbsugAACAAJ&dq=intitle:relativistic+figures+of+equilibrium+inauthor:ansorg&hl=&cd=1&source=gbs_api}

\bibitem{Schenk:1999jv}
Schenk A~K, Shapiro S~L and Teukolsky S~A 1999 {\em The Astrophysical
  Journal\/} {\bf 521} 310--318
  \urlprefix\url{http://stacks.iop.org/0004-637X/521/i=1/a=310}

\bibitem{Shapiro:1991kg}
Shapiro S~L and Teukolsky S~A 1991 {\em Phys. Rev. Lett.\/} {\bf 66} 994--997
  \urlprefix\url{http://link.aps.org/doi/10.1103/PhysRevLett.66.994}

\bibitem{Fricke:1952vb}
Fricke W 1952 {\em Astronomische Nachrichten. News in Astronomy and
  Astrophysics\/} {\bf 280} 193--216
  \urlprefix\url{http://dx.doi.org/10.1002/asna.19522800502}

\bibitem{Toomre:1982aa}
Toomre A 1982 {\em The Astrophysical Journal\/} {\bf 259} 535--543
  \urlprefix\url{http://adsabs.harvard.edu/full/1982ApJ...259..535T}

\bibitem{Evans:vc}
Evans N~W 1993 {\em Monthly Notices of the Royal Astronomical Society\/} {\bf
  260} 191--201
  \urlprefix\url{http://mnras.oxfordjournals.org/content/260/1/191.short}

\bibitem{Binney:1981aa}
Binney J 1981 {\em Monthly Notices of the Royal Astronomical Society\/} {\bf
  196} 455--467
  \urlprefix\url{http://mnras.oxfordjournals.org/content/196/3/455.abstract}

\bibitem{Firt:2009hv}
Fi{\v r}t R, Rein G and Seehafer M 2009 {\em Communications in Mathematical
  Physics\/} {\bf 291} 225--255
  \urlprefix\url{http://link.springer.com/10.1007/s00220-009-0872-7}

\bibitem{HOAG:1950vn}
HOAG A~A 1950 {\em Astronomical Journal\/} {\bf 55} 170--170
  \urlprefix\url{http://dx.doi.org/10.1086/106427}

\end{thebibliography}
}

\end{document}